# The Three-Dimensionality of Momentum as a Property of Photons

*Xiaoshuang Shen*

School of Physical Science and Technology, Yangzhou University, Yangzhou, 225002, P. R. China

E-mail: xsshen@yzu.edu.cn

**Abstract**

In the theory of modern physics, such as in relativity and quantum mechanics, the three-dimensionality of space is introduced as a presupposed fact. The three-dimensionality of particle motion, that is, the three-dimensionality of particle momentum, is a result of this fact. In this paper, I make two assumptions on the spin properties of photons, and then deduce a relation between the direction of photon momentum and the spin plane to which the photon spin state belongs. Based on this relation, the three-dimensionality of photon momentum is taken as the photon's own property. Moreover, this relation allows us to better understand some of the well-known theoretical results, such as the change rules of the spin states of spin-1 particles under space rotation as well as the relation between the momentum direction of the classical electromagnetic wave and the directions of the electric and magnetic fields.

At least since ancient Greece, people realized that the space in which we live is three-dimensional.[1] However, To the ancient question of why space is three-dimensional, physicists have so far failed to give a straightforward explanation.[2] In the usual physical description, it is always assumed that the three-dimensionality is a property of the space itself, and the three-dimensionality of momentum is a natural result of this property. However, if we look closely at the experiences which make us think that the space is three-dimensional, we will find that they are directly or indirectly related to the momentum of photons. The orientation of the celestial bodies is determined by the light from them; the different orientations correspond to the different momentum directions of the photons. The three-dimensionality of the spatial structure of the objects around us, including various experimental instruments, is also known from the light. The momentum of the light from the different parts of these objects to the reference point has a three-dimensional structure. These considerations suggest a logical possibility that the three-dimensionality of the photon momentum is a property of the photon itself, and the three-dimensionality of the space is a result of this property of photons. Henry Poincaré once proposed that "…the properties of space are merely those of the measuring instruments", and in addition, "…a ray of light is also one of our instruments".[3] The special relativity provides an example of support for the view of Poincaré. The propagation velocity of light in vacuum is independent of the choice of the inertia system. Accordingly, the space and time coordinates in different inertial systems satisfy the Lorentz transformation.

As a kind of massless particles, all photons have momentum due to the fact that they cannot stand still in any inertial system. This is different from non-zero mass particles. The latter's momentum can be zero in a particular reference frame. Thus, it is difficult to regard the characteristics of the momentum of these particles as the nature of these particles themselves. As a kind of spin-1 particles, the spin space of photons is expected to be three-dimensional. However, the description of the spin state of a photon with a given momentum direction $\hat{\mathbf{k}}$ requires only two base states, such as the right-handed polarization state $|S_{\hat{\mathbf{k}}} = +\hbar\rangle$ and the left-handed polarization state $|S_{\hat{\mathbf{k}}} = -\hbar\rangle$. We will show below that if we make two assumptions about the spin properties of the photons based on the well-known properties of photons, the momentum of the photons will have those features we are familiar with, such as the three-dimensionality, etc.

The two assumptions are as follows:

**Assumption Ⅰ (Correspondence Assumption)** The spin state of a photon always belongs to a spin plane, which determines the momentum direction of the photon.[4]

**Assumption Ⅱ (Symmetry Assumption)** All the possible spin planes to which the spin state of a photon may belong can be defined by the two following orthonormal spin states:

$$|S_{\hat{\mathbf{k}}} = +\hbar\rangle = \frac{1}{2}(1+\cos\beta)e^{-i\alpha}|a\rangle + \frac{1}{\sqrt{2}}\sin\beta|b\rangle + \frac{1}{2}(1-\cos\beta)e^{i\alpha}|c\rangle, \tag{1a}$$

$$|S_{\hat{\mathbf{k}}} = -\hbar\rangle = \frac{1}{2}(1-\cos\beta)e^{-i\alpha}|a\rangle - \frac{1}{\sqrt{2}}\sin\beta|b\rangle + \frac{1}{2}(1+\cos\beta)e^{i\alpha}|c\rangle, \tag{1b}$$

where $|a\rangle$, $|b\rangle$ and $|c\rangle$ are the three spin base states of the spin-1 particles, and α and β are two real numbers.

The **Assumption Ⅰ** shows that there is a correspondence between the spin plane and the momentum direction of a photon. This correspondence can explain why the description of the spin state of a photon with a given momentum direction $\hat{\mathbf{k}}$ only requires two base states. We write this correspondence as

$$f(|S_{\hat{\mathbf{k}}} = +\hbar\rangle, |S_{\hat{\mathbf{k}}} = -\hbar\rangle) = \hat{\mathbf{k}}. \tag{2}$$

The above equation means that if the spin state of a photon belongs to the spin plane determined by $|S_{\hat{\mathbf{k}}} = +\hbar\rangle$ and $|S_{\hat{\mathbf{k}}} = -\hbar\rangle$, the photon will have the momentum along the direction $\hat{\mathbf{k}}$ through some mechanism that we still do not know. For photons, when the momentum direction is reversed, the original left- and right-handed polarization states are interchanged. Thus, the mapping function *f* has the property

$$f(|S_{\hat{\mathbf{k}}} = -\hbar\rangle, |S_{\hat{\mathbf{k}}} = +\hbar\rangle) = -\hat{\mathbf{k}} = f(|S_{\hat{\mathbf{k}}} = +\hbar\rangle, |S_{\hat{\mathbf{k}}} = -\hbar\rangle). \tag{3}$$

That is to say, the direction of the photon momentum corresponds to the "directional" spin plane.

For a photon whose momentum is in the direction of $+\hat{\mathbf{k}}$ or $-\hat{\mathbf{k}}$, the base states of its spin can also be chosen as other two orthonormal states which are linear combination of $|S_{\hat{\mathbf{k}}} = +\hbar\rangle$ and $|S_{\hat{\mathbf{k}}} = -\hbar\rangle$:

$$|1\rangle = C_{11}|S_{\hat{\mathbf{k}}} = +\hbar\rangle + C_{12}|S_{\hat{\mathbf{k}}} = -\hbar\rangle, \qquad (4a)$$

$$|2\rangle = C_{21}|S_{\hat{\mathbf{k}}} = +\hbar\rangle + C_{22}|S_{\hat{\mathbf{k}}} = -\hbar\rangle. \qquad (4b)$$

The orthonormality of $|1\rangle$ and $|2\rangle$ requires that $C_{ij}(i,j=1,2)$ are the matrix elements of a $U(2)$ matrix $C$. Since the spin plane determined by the base states $|1\rangle$ and $|2\rangle$ corresponds to the $+\hat{\mathbf{k}}$ or $-\hat{\mathbf{k}}$ direction, we have

$$\begin{aligned} f(|1\rangle,|2\rangle) &= f(C_{11}|S_{\hat{\mathbf{k}}} = +\hbar\rangle + C_{12}|S_{\hat{\mathbf{k}}} = -\hbar\rangle, C_{21}|S_{\hat{\mathbf{k}}} = +\hbar\rangle + C_{22}|S_{\hat{\mathbf{k}}} = -\hbar\rangle) \\ &= g(C_{11},C_{12},C_{21},C_{22})\hat{\mathbf{k}}, \end{aligned} \qquad (5)$$

where $g(C_{11},C_{12},C_{21},C_{22}) = \pm 1$. Let us analyze the function $g(C_{11},C_{12},C_{21},C_{22})$. When $C_{11} = C_{22} = 1$ and $C_{12} = C_{21} = 0$, we have

$$g(1,0,0,1) = 1. \qquad (6a)$$

When $C_{11} = C_{22} = 0$ and $C_{12} = C_{21} = 1$, we have

$$g(0,1,1,0) = -1. \qquad (6b)$$

In addition, the base states $|1\rangle$ and $|2\rangle$ should satisfy $f(|1\rangle,|2\rangle) = -f(|2\rangle,|1\rangle)$, so we have

$$g(C_{11},C_{12},C_{21},C_{22}) = -g(C_{21},C_{22},C_{11},C_{12}). \qquad (6c)$$

Obviously, the function

$$g(C_{11},C_{12},C_{21},C_{22}) = C_{11}C_{22} - C_{12}C_{21} \qquad (7)$$

can satisfy all the Eqs. (6a-c). The right side of Eq. (7) is the determinant of the matrix $C$. In choosing the base states, if we multiply all the base states by a common phase factor $e^{i\theta}$ ($\theta$ is a real number), it does not have any effect on the description of the physical states. Therefore, we can always absorb the appropriate phase factors into the definition of the base states so that the determinant of matrix $C$ is real, i.e. $\det C = \pm 1$.[5] Equations (6a, b) are the corresponding special cases. When the two rows of matrix $C$ are interchanged, the value of its determinant changes the sign. This corresponds to Eq. (6c). Combining Eqs. (2), (5) and (7), we have

$$\begin{aligned} &f(C_{11}|S_{\hat{\mathbf{k}}} = +\hbar\rangle + C_{12}|S_{\hat{\mathbf{k}}} = -\hbar\rangle, C_{21}|S_{\hat{\mathbf{k}}} = +\hbar\rangle + C_{22}|S_{\hat{\mathbf{k}}} = -\hbar\rangle) \\ &= (C_{11}C_{22} - C_{21}C_{21})\hat{\mathbf{k}} = (C_{11}C_{22} - C_{12}C_{21})f(|S_{\hat{\mathbf{k}}} = +\hbar\rangle,|S_{\hat{\mathbf{k}}} = -\hbar\rangle). \end{aligned} \qquad (8)$$

From the above analysis, we always require that any two sets of base states of a same spin plane are associated with a $SU(2)$ matrix.

Let us consider the **Assumption Ⅱ**. Because $|S_{\hat{\mathbf{k}}} = +\hbar\rangle_{\alpha=\alpha_0+2\pi,\beta=\beta_0} = |S_{\hat{\mathbf{k}}} = +\hbar\rangle_{\alpha=\alpha_0,\beta=\beta_0}$ and

$\left|S_{\hat{\mathbf{k}}}=-\hbar\right\rangle_{\alpha=\alpha_0+2\pi,\beta=\beta_0} = \left|S_{\hat{\mathbf{k}}}=-\hbar\right\rangle_{\alpha=\alpha_0,\beta=\beta_0}$, we limit the range of α to $[0,2\pi)$. For β, we similarly have $\left|S_{\hat{\mathbf{k}}}=+\hbar\right\rangle_{\alpha=\alpha_0,\beta=\beta_0+2\pi} = \left|S_{\hat{\mathbf{k}}}=+\hbar\right\rangle_{\alpha=\alpha_0,\beta=\beta_0}$ and $\left|S_{\hat{\mathbf{k}}}=-\hbar\right\rangle_{\alpha=\alpha_0,\beta=\beta_0+2\pi} = \left|S_{\hat{\mathbf{k}}}=-\hbar\right\rangle_{\alpha=\alpha_0,\beta=\beta_0}$. And because of $\left|S_{\hat{\mathbf{k}}}=+\hbar\right\rangle_{\alpha=\alpha_0,\beta=\beta_0+\pi} = -\left|S_{\hat{\mathbf{k}}}=+\hbar\right\rangle_{\alpha=\alpha_0+\pi,\beta=\pi-\beta_0}$ and $\left|S_{\hat{\mathbf{k}}}=-\hbar\right\rangle_{\alpha=\alpha_0,\beta=\beta_0+\pi} = -\left|S_{\hat{\mathbf{k}}}=-\hbar\right\rangle_{\alpha=\alpha_0+\pi,\beta=\pi-\beta_0}$, we limit the range of β to $[0,\pi]$.

Define three new base states:

$$|x\rangle = -\frac{1}{\sqrt{2}}e^{i\pi/4}(|a\rangle-|c\rangle), \quad |y\rangle = \frac{i}{\sqrt{2}}e^{i\pi/4}(|a\rangle-|c\rangle), \quad |z\rangle = e^{i\pi/4}|b\rangle. \qquad (9)$$

With these three base states, Eqs. (1a, b) can be rewritten as

$$\left|S_{\hat{\mathbf{k}}}=+\hbar\right\rangle = \frac{1}{\sqrt{2}}(i\sin\alpha - \cos\alpha\cos\beta)e^{-i\pi/4}|x\rangle \qquad (10a)$$
$$+\frac{1}{\sqrt{2}}(-i\cos\alpha - \sin\alpha\cos\beta)e^{-i\pi/4}|y\rangle + \frac{1}{\sqrt{2}}\sin\beta e^{-i\pi/4}|z\rangle,$$

$$\left|S_{\hat{\mathbf{k}}}=-\hbar\right\rangle = \frac{1}{\sqrt{2}}(i\sin\alpha + \cos\alpha\cos\beta)e^{-i\pi/4}|x\rangle$$
$$+\frac{1}{\sqrt{2}}(-i\cos\alpha + \sin\alpha\cos\beta)e^{-i\pi/4}|y\rangle - \frac{1}{\sqrt{2}}\sin\beta e^{-i\pi/4}|z\rangle.$$

(10b)

Substituting Eqs.(10a, b) into Eq.(2), we have

$$\hat{\mathbf{k}} = f(\frac{1}{\sqrt{2}}(i\sin\alpha - \cos\alpha\cos\beta)e^{-i\pi/4}|x\rangle + \frac{1}{\sqrt{2}}(-i\cos\alpha - \sin\alpha\cos\beta)e^{-i\pi/4}|y\rangle + \frac{1}{\sqrt{2}}\sin\beta e^{-i\pi/4}|z\rangle,$$
$$\frac{1}{\sqrt{2}}(i\sin\alpha + \cos\alpha\cos\beta)e^{-i\pi/4}|x\rangle + \frac{1}{\sqrt{2}}(-i\cos\alpha + \sin\alpha\cos\beta)e^{-i\pi/4}|y\rangle - \frac{1}{\sqrt{2}}\sin\beta e^{-i\pi/4}|z\rangle).$$

(11)

Let us analyze the structure of the photon momentum space according to Eq. (11). For any two spin planes in the photon spin space, if there is a spin state in one of the spin planes orthogonal to all the spin states in the other spin plane, we say that the two spin planes orthogonal. From the base states $|x\rangle$, $|y\rangle$ and $|z\rangle$, we can obtain three mutually orthogonal spin planes: $(|x\rangle,|y\rangle)$, $(|y\rangle,|z\rangle)$ and $(|z\rangle,|x\rangle)$. With respect to these three spin planes, we can prove the following three equations:

$$f(|x\rangle,|y\rangle) = f(|S_{\hat{\mathbf{z}}}=+\hbar\rangle,|S_{\hat{\mathbf{z}}}=-\hbar\rangle), \qquad (12a)$$

$$f(|y\rangle,|z\rangle) = f(|S_{\hat{\mathbf{x}}}=+\hbar\rangle,|S_{\hat{\mathbf{x}}}=-\hbar\rangle), \qquad (12b)$$

$$f(|z\rangle,|x\rangle) = f(|S_{\hat{\mathbf{y}}}=+\hbar\rangle,|S_{\hat{\mathbf{y}}}=-\hbar\rangle), \qquad (12c)$$

where,

$$\left|S_{\hat{\mathbf{x}}}=+\hbar\right\rangle \equiv \left|S_{\hat{\mathbf{k}}}=+\hbar\right\rangle_{\alpha=0,\beta=\pi/2} = -\frac{i}{\sqrt{2}}e^{-i\pi/4}|y\rangle + \frac{1}{\sqrt{2}}e^{-i\pi/4}|z\rangle, \qquad (13a)$$

$$\left|S_{\hat{x}}=-\hbar\right\rangle \equiv \left|S_{\hat{k}}=-\hbar\right\rangle_{\alpha=0,\beta=\pi/2} = -\frac{i}{\sqrt{2}}e^{-i\pi/4}|y\rangle - \frac{1}{\sqrt{2}}e^{-i\pi/4}|z\rangle, \quad (13b)$$

$$\left|S_{\hat{y}}=+\hbar\right\rangle \equiv \left|S_{\hat{k}}=+\hbar\right\rangle_{\alpha=\pi/2,\beta=\pi/2} = \frac{i}{\sqrt{2}}e^{-i\pi/4}|x\rangle + \frac{1}{\sqrt{2}}e^{-i\pi/4}|z\rangle, \quad (13c)$$

$$\left|S_{\hat{y}}=-\hbar\right\rangle \equiv \left|S_{\hat{k}}=-\hbar\right\rangle_{\alpha=\pi/2,\beta=\pi/2} = \frac{i}{\sqrt{2}}e^{-i\pi/4}|x\rangle - \frac{1}{\sqrt{2}}e^{-i\pi/4}|z\rangle, \quad (13d)$$

$$\left|S_{\hat{z}}=+\hbar\right\rangle \equiv \left|S_{\hat{k}}=+\hbar\right\rangle_{\alpha=0,\beta=0} = -\frac{1}{\sqrt{2}}e^{-i\pi/4}|x\rangle - \frac{i}{\sqrt{2}}e^{-i\pi/4}|y\rangle, \quad (13e)$$

$$\left|S_{\hat{z}}=-\hbar\right\rangle \equiv \left|S_{\hat{k}}=-\hbar\right\rangle_{\alpha=0,\beta=0} = \frac{1}{\sqrt{2}}e^{-i\pi/4}|x\rangle - \frac{i}{\sqrt{2}}e^{-i\pi/4}|y\rangle. \quad (13f)$$

We give the proof of Eq. (12a) below, and the proofs of Eqs. (12b, c) are given in SI.[6] According to Eqs. (13e, f) and (8), we have

$$\begin{aligned}
&f(\left|S_{\hat{z}}=+\hbar\right\rangle, \left|S_{\hat{z}}=-\hbar\right\rangle) \\
&= f(-\frac{1}{\sqrt{2}}e^{-i\pi/4}|x\rangle - \frac{i}{\sqrt{2}}e^{-i\pi/4}|y\rangle, \frac{1}{\sqrt{2}}e^{-i\pi/4}|x\rangle - \frac{i}{\sqrt{2}}e^{-i\pi/4}|y\rangle) \\
&= [(-\frac{1}{\sqrt{2}}e^{-i\pi/4})(-\frac{i}{\sqrt{2}}e^{-i\pi/4}) - (-\frac{i}{\sqrt{2}}e^{-i\pi/4})(\frac{1}{\sqrt{2}}e^{-i\pi/4})]f(|x\rangle,|y\rangle) \\
&= f(|x\rangle,|y\rangle).
\end{aligned} \quad (14)$$

The three spin planes appearing on the right side of the Eqs. (12a-c) are obtained from the two base states defined by Eqs. (10a, 10b) with different given values of α and β, so that they all correspond to the directions in which the actual photon momentum can be. Denoting these three directions of the photon momentum as $\hat{\mathbf{k}}_z$, $\hat{\mathbf{k}}_x$ and $\hat{\mathbf{k}}_y$, respectively, we have

$$f(|x\rangle,|y\rangle) = \hat{\mathbf{k}}_z, \quad f(|y\rangle,|z\rangle) = \hat{\mathbf{k}}_x, \quad f(|z\rangle,|x\rangle) = \hat{\mathbf{k}}_y. \quad (15)$$

To simplify the expression, we write the Eqs. (10a, b) as follows:

$$\left|S_{\hat{k}}=+\hbar\right\rangle = d_{11}|x\rangle + d_{12}|y\rangle + d_{13}|z\rangle, \quad (16a)$$

$$\left|S_{\hat{k}}=-\hbar\right\rangle = d_{21}|x\rangle + d_{22}|y\rangle + d_{23}|z\rangle, \quad (16b)$$

where $d_{ij}(i=1,2; j=1,2,3)$ represent the parameter expressions before the corresponding base states in Eqs. (10a, b).[7] Thus, Eq. (11) can be written as

$$\hat{\mathbf{k}} = f(d_{11}|x\rangle + d_{12}|y\rangle + d_{13}|z\rangle, d_{21}|x\rangle + d_{22}|y\rangle + d_{23}|z\rangle) \quad (17)$$

Let's analyze the relationship between $\hat{\mathbf{k}}$ and the three directions $\hat{\mathbf{k}}_z$, $\hat{\mathbf{k}}_x$ and $\hat{\mathbf{k}}_y$. When $d_{11} = d_{21} = 0$, according to Eqs. (8) and (15), we have

$$\begin{aligned}
\hat{\mathbf{k}} &= f(d_{12}|y\rangle + d_{13}|z\rangle, d_{22}|y\rangle + d_{23}|z\rangle) \\
&= (d_{12}d_{23} - d_{13}d_{22})f(|y\rangle,|z\rangle) = (d_{12}d_{23} - d_{13}d_{22})\hat{\mathbf{k}}_x.
\end{aligned} \quad (18a)$$

Similarly, when $d_{12} = d_{22} = 0$ and $d_{13} = d_{23} = 0$, we have

$$\hat{\mathbf{k}} = f(d_{11}|x\rangle + d_{13}|z\rangle, d_{21}|x\rangle + d_{23}|z\rangle) \qquad (18b)$$
$$= (d_{13}d_{21} - d_{11}d_{23})f(|x\rangle,|z\rangle) = (d_{13}d_{21} - d_{11}d_{23})\hat{\mathbf{k}}_{\mathbf{y}},$$

$$\hat{\mathbf{k}} = f(d_{11}|x\rangle + d_{12}|y\rangle, d_{21}|x\rangle + d_{22}|y\rangle) \qquad (18c)$$
$$= (d_{11}d_{22} - d_{12}d_{21})f(|x\rangle,|y\rangle) = (d_{11}d_{22} - d_{12}d_{21})\hat{\mathbf{k}}_{\mathbf{z}},$$

respectively. Suppose the relation between $\hat{\mathbf{k}}$ and the three directions $\hat{\mathbf{k}}_{\mathbf{z}}$, $\hat{\mathbf{k}}_{\mathbf{x}}$ and $\hat{\mathbf{k}}_{\mathbf{y}}$ is linear, and thus we have

$$\begin{aligned}\hat{\mathbf{k}} = &\, g_1(d_{11},d_{12},d_{13},d_{21},d_{22},d_{23})(d_{12}d_{23} - d_{13}d_{22})\hat{\mathbf{k}}_{\mathbf{x}} \\ &+ g_2(d_{11},d_{12},d_{13},d_{21},d_{22},d_{23})(d_{13}d_{21} - d_{11}d_{23})\hat{\mathbf{k}}_{\mathbf{y}} \\ &+ g_3(d_{11},d_{12},d_{13},d_{21},d_{22},d_{23})(d_{11}d_{22} - d_{12}d_{21})\hat{\mathbf{k}}_{\mathbf{z}}.\end{aligned} \qquad (19)$$

According to Eqs. (18a-c), we have

$$g_1(0,d_{12},d_{13},0,d_{22},d_{23}) = g_2(d_{11},0,d_{13},d_{21},0,d_{23}) = g_3(d_{11},d_{12},0,d_{21},d_{22},0) = 1. \qquad (20)$$

In addition, according to Eq. (3), we have

$$\begin{aligned}&f(d_{11}|x\rangle + d_{12}|y\rangle + d_{13}|z\rangle, d_{21}|x\rangle + d_{22}|y\rangle + d_{23}|z\rangle) \\ &= g_1(d_{11},d_{12},d_{13},d_{21},d_{22},d_{23})(d_{12}d_{23} - d_{13}d_{22})\hat{\mathbf{k}}_{\mathbf{x}} \\ &+ g_2(d_{11},d_{12},d_{13},d_{21},d_{22},d_{23})(d_{13}d_{21} - d_{11}d_{23})\hat{\mathbf{k}}_{\mathbf{y}} \\ &+ g_3(d_{11},d_{12},d_{13},d_{21},d_{22},d_{23})(d_{11}d_{22} - d_{12}d_{21})\hat{\mathbf{k}}_{\mathbf{z}} \\ &= -f(d_{21}|x\rangle + d_{22}|y\rangle + d_{23}|z\rangle, d_{11}|x\rangle + d_{12}|y\rangle + d_{13}|z\rangle) \\ &= g_1(d_{21},d_{22},d_{23},d_{11},d_{12},d_{13})(d_{12}d_{23} - d_{13}d_{22})\hat{\mathbf{k}}_{\mathbf{x}} \\ &+ g_2(d_{21},d_{22},d_{23},d_{11},d_{12},d_{13})(d_{13}d_{21} - d_{11}d_{23})\hat{\mathbf{k}}_{\mathbf{y}} \\ &+ g_3(d_{21},d_{22},d_{23},d_{11},d_{12},d_{13})(d_{11}d_{22} - d_{12}d_{21})\hat{\mathbf{k}}_{\mathbf{z}}.\end{aligned} \qquad (21)$$

Obviously, taking

$$g_i(d_{11},d_{12},d_{13},d_{21},d_{22},d_{23}) \equiv 1, \quad (i=1,2,3), \qquad (22)$$

can satisfy Eqs. (20) and (21). Substituting Eq. (22) into Eq. (19) and using Eqs. (17) and (15), we have

$$\begin{aligned}\hat{\mathbf{k}} &= (d_{12}d_{23} - d_{13}d_{22})\hat{\mathbf{k}}_{\mathbf{x}} + (d_{13}d_{21} - d_{11}d_{23})\hat{\mathbf{k}}_{\mathbf{y}} + (d_{11}d_{22} - d_{12}d_{21})\hat{\mathbf{k}}_{\mathbf{z}} \\ &= f(d_{11}|x\rangle + d_{12}|y\rangle + d_{13}|z\rangle, d_{21}|x\rangle + d_{22}|y\rangle + d_{23}|z\rangle) \\ &= (d_{12}d_{23} - d_{13}d_{22})\, f(|y\rangle,|z\rangle) + (d_{13}d_{21} - d_{11}d_{23})\, f(|z\rangle,|x\rangle) + (d_{11}d_{22} - d_{12}d_{21})\, f(|x\rangle,|y\rangle).\end{aligned} \qquad (23)$$

Applying Eq. (23) to Eq. (11), we obtain

$$\hat{\mathbf{k}} = f(\frac{1}{\sqrt{2}}(i\sin\alpha - \cos\alpha\cos\beta)e^{-i\pi/4}|x\rangle + \frac{1}{\sqrt{2}}(-i\cos\alpha - \sin\alpha\cos\beta)e^{-i\pi/4}|y\rangle + \frac{1}{\sqrt{2}}\sin\beta e^{-i\pi/4}|z\rangle,$$

$$\frac{1}{\sqrt{2}}(i\sin\alpha + \cos\alpha\cos\beta)e^{-i\pi/4}|x\rangle + \frac{1}{\sqrt{2}}(-i\cos\alpha + \sin\alpha\cos\beta)e^{-i\pi/4}|y\rangle - \frac{1}{\sqrt{2}}\sin\beta e^{-i\pi/4}|z\rangle)$$

$$= [\frac{1}{\sqrt{2}}(-i\cos\alpha - \sin\alpha\cos\beta)e^{-i\pi/4}(-\frac{1}{\sqrt{2}})\sin\beta e^{-i\pi/4}$$

$$-\frac{1}{\sqrt{2}}\sin\beta e^{-i\pi/4}\frac{1}{\sqrt{2}}(-i\cos\alpha + \sin\alpha\cos\beta)e^{-i\pi/4}]f(|y\rangle,|z\rangle)$$

$$+[\frac{1}{\sqrt{2}}\sin\beta e^{-i\pi/4}\frac{1}{\sqrt{2}}(i\sin\alpha + \cos\alpha\cos\beta)e^{-i\pi/4}$$

$$-\frac{1}{\sqrt{2}}(i\sin\alpha - \cos\alpha\cos\beta)e^{-i\pi/4}(-\frac{1}{\sqrt{2}})\sin\beta e^{-i\pi/4}]f(|z\rangle,|x\rangle)$$

$$+[\frac{1}{\sqrt{2}}(i\sin\alpha - \cos\alpha\cos\beta)e^{-i\pi/4}\frac{1}{\sqrt{2}}(-i\cos\alpha + \sin\alpha\cos\beta)e^{-i\pi/4}$$

$$-\frac{1}{\sqrt{2}}(-i\cos\alpha - \sin\alpha\cos\beta)e^{-i\pi/4}\frac{1}{\sqrt{2}}(i\sin\alpha + \cos\alpha\cos\beta)e^{-i\pi/4}]f(|x\rangle,|y\rangle)$$

$$= \sin\beta\cos\alpha f(|y\rangle,|z\rangle) + \sin\beta\sin\alpha f(|z\rangle,|x\rangle) + \cos\beta f(|x\rangle,|y\rangle)$$

$$= \sin\beta\cos\alpha\hat{\mathbf{k}}_{\mathbf{x}} + \sin\beta\sin\alpha\hat{\mathbf{k}}_{\mathbf{y}} + \cos\beta\hat{\mathbf{k}}_{\mathbf{z}}.$$

(24)

If we compare the selected momentum directions $\hat{\mathbf{k}}_{\mathbf{x}}$, $\hat{\mathbf{k}}_{\mathbf{y}}$ and $\hat{\mathbf{k}}_{\mathbf{z}}$ to the three unit direction vectors $\hat{\mathbf{e}}_{\mathbf{x}}$, $\hat{\mathbf{e}}_{\mathbf{y}}$ and $\hat{\mathbf{e}}_{\mathbf{z}}$ in the Cartesian coordinate system, the three components in Eq. (24) are exactly the three Cartesian components of a unit vector represented by the spherical coordinates α and β. Thus, the parameters α and β, which come from the Eqs. (1a, b), are azimuth and polar angles in the spherical coordinate system, respectively.

We have derived the expressions of the correspondence between the spin plane of a photon and the direction of the photon momentum based on the **Assumption** Ⅰ and **Assumption** Ⅱ. It can be shown that this correspondence is one-to-one. That is to say, the same spin plane corresponds to the same momentum direction; the same momentum direction also corresponds to the same spin plane.[8] From the structure of the momentum space represented by Eq. (24), it can be seen that the **Assumption** Ⅱ is equivalent to assuming that all the spin planes which the photon spin state may belong can be obtained through all the *SO*(3) rotations from one of these spin planes. This is the reason why the **Assumption** Ⅱ is referred to as the **Symmetry Assumption**. Based on this one-to-one correspondence we can prove that it is impossible to decompose all the photon momentums into two selected directions.[9]

The momentum of the photon is decomposed into three directions corresponding to three special spin planes (Eqs. (12a-c)). The momentum of the photons can also be decomposed into the directions corresponding to any three selected spin planes, as long as the three directions are not coplanar. Take three spin planes $(|S_{\hat{\mathbf{k}}} = +\hbar\rangle_{\alpha=\alpha_1,\beta=\beta_1}, |S_{\hat{\mathbf{k}}} = -\hbar\rangle_{\alpha=\alpha_1,\beta=\beta_1})$, $(|S_{\hat{\mathbf{k}}} = +\hbar\rangle_{\alpha=\alpha_2,\beta=\beta_2}, |S_{\hat{\mathbf{k}}} = -\hbar\rangle_{\alpha=\alpha_2,\beta=\beta_2})$ and $(|S_{\hat{\mathbf{k}}} = +\hbar\rangle_{\alpha=\alpha_3,\beta=\beta_3}, |S_{\hat{\mathbf{k}}} = -\hbar\rangle_{\alpha=\alpha_3,\beta=\beta_3})$. The corresponding momentum directions are represented by $\hat{\mathbf{k}}_1$, $\hat{\mathbf{k}}_2$ and $\hat{\mathbf{k}}_3$ respectively, and thus

we have

$$\hat{\mathbf{k}} = g'_1(\alpha,\beta)\hat{\mathbf{k}}_1 + g'(\alpha,\beta)_2 \hat{\mathbf{k}}_2 + g'_3(\alpha,\beta)\hat{\mathbf{k}}_3 .\qquad(25)$$

We decompose $\hat{\mathbf{k}}$, $\hat{\mathbf{k}}_1$, $\hat{\mathbf{k}}_2$ and $\hat{\mathbf{k}}_3$ according to Eq. (24), and thus have

$$\begin{aligned}&\sin\beta\cos\alpha\,\hat{\mathbf{k}}_x + \sin\beta\sin\alpha\,\hat{\mathbf{k}}_y + \cos\beta\,\hat{\mathbf{k}}_z\\&= [g'_1(\alpha,\beta)\sin\beta_1\cos\alpha_1 + g'_2(\alpha,\beta)\sin\beta_2\cos\alpha_2 + g'_3(\alpha,\beta)\sin\beta_3\cos\alpha_3]\hat{\mathbf{k}}_x\\&+ [g'_1(\alpha,\beta)\sin\beta_1\sin\alpha_1 + g'_2(\alpha,\beta)\sin\beta_2\sin\alpha_2 + g'_3(\alpha,\beta)\sin\beta_3\sin\alpha_3]\hat{\mathbf{k}}_y\\&+ [g'_1(\alpha,\beta)\cos\beta_1 + g'_2(\alpha,\beta)\cos\beta_2 + g'_3(\alpha,\beta)\cos\beta_3]\hat{\mathbf{k}}_z.\end{aligned}\qquad(26)$$

Because the correspondence between the spin plane of the photon and the direction of the photon momentum is one-to-one, we have

$$\sin\beta\cos\alpha = g'_1(\alpha,\beta)\sin\beta_1\cos\alpha_1 + g'_2(\alpha,\beta)\sin\beta_2\cos\alpha_2 + g'_3(\alpha,\beta)\sin\beta_3\cos\alpha_3 ,\qquad(27a)$$

$$\sin\beta\sin\alpha = g'_1(\alpha,\beta)\sin\beta_1\sin\alpha_1 + g'_2(\alpha,\beta)\sin\beta_2\sin\alpha_2 + g'_3(\alpha,\beta)\sin\beta_3\sin\alpha_3 ,\qquad(27b)$$

$$\cos\beta = g'_1(\alpha,\beta)\cos\beta_1 + g'_2(\alpha,\beta)\cos\beta_2 + g'_3(\alpha,\beta)\cos\beta_3 .\qquad(27c)$$

When the determinant of the matrix

$$B = \begin{pmatrix} \sin\beta_1\cos\alpha_1 & \sin\beta_2\cos\alpha_2 & \sin\beta_3\cos\alpha_3 \\ \sin\beta_1\sin\alpha_1 & \sin\beta_2\sin\alpha_2 & \sin\beta_3\sin\alpha_3 \\ \cos\beta_1 & \cos\beta_2 & \cos\beta_3 \end{pmatrix}\qquad(28)$$

is not equal to 0 (i.e. $\det B \neq 0$), we can obtain

$$g'_i(\alpha,\beta) = \det B_i / \det B, (i=1,2,3) ,\qquad(29)$$

where the matrix $B_i$ is a matrix obtained through substituting the matrix elements of the i-th column of the matrix B by the corresponding expressions on the left side of the Eqs. (27a-c).[10] It can be shown that the condition $\det B \neq 0$ exactly requires that $\hat{\mathbf{k}}_1$, $\hat{\mathbf{k}}_2$ and $\hat{\mathbf{k}}_3$ are not coplanar.[11] According to the Eqs. (27a-c), the relation expressed by Eq. (25) is exactly the relation between a nomal unit space vector and its three components. That is, $\hat{\mathbf{k}}$, $\hat{\mathbf{k}}_1$, $\hat{\mathbf{k}}_2$ and $\hat{\mathbf{k}}_3$ correspond to the diagonal line and the three sides of a parallelepiped that cross a same vertex, respectively. It can also be shown that if $\hat{\mathbf{k}}_1$, $\hat{\mathbf{k}}_2$ and $\hat{\mathbf{k}}_3$ are three mutually perpendicular directions then we have

$$\sum_{i=1}^{3} g'^{2}_i(\alpha,\beta) = 1 ,\qquad(30)$$

and vice versa.[12] Once again, this is consistent with the relation between a space vector and its components. It can also be proved that for photons, if two momentum directions are perpendicular to each other, the corresponding spin planes are orthogonal, and vice versa.[13]

The base states $|x\rangle$, $|y\rangle$ and $|z\rangle$ determine the three momentum directions of photons, $\hat{\mathbf{k}}_1$,

$\hat{\mathbf{k}}_2$ and $\hat{\mathbf{k}}_3$, by Eq. (15). We take other three mutually perpendicular directions $\hat{\mathbf{k}}_{x'}$, $\hat{\mathbf{k}}_{y'}$ and $\hat{\mathbf{k}}_{z'}$ in the momentum space of the photon, satisfying

$$\begin{pmatrix} \hat{\mathbf{k}}_{x'} \\ \hat{\mathbf{k}}_{y'} \\ \hat{\mathbf{k}}_{z'} \end{pmatrix} = M \begin{pmatrix} \hat{\mathbf{k}}_x \\ \hat{\mathbf{k}}_y \\ \hat{\mathbf{k}}_z \end{pmatrix}, \qquad (31)$$

where, $M$ is a 3 × 3 real orthogonal matrix, and $\det M = 1$. Take a new set of base states $|x'\rangle$, $|y'\rangle$ and $|z'\rangle$, satisfying

$$\begin{pmatrix} |x'\rangle \\ |y'\rangle \\ |z'\rangle \end{pmatrix} = N \begin{pmatrix} |x\rangle \\ |y\rangle \\ |z\rangle \end{pmatrix}, \qquad (32)$$

where, $N$ also is a 3 × 3 real orthogonal matrix, and $\det N = 1$. If

$$\hat{\mathbf{k}}_{x'} = f(|y'\rangle, |z'\rangle), \quad \hat{\mathbf{k}}_{y'} = f(|z'\rangle, |x'\rangle), \quad \hat{\mathbf{k}}_{z'} = f(|x'\rangle, |y'\rangle) \qquad (33)$$

hold, we can prove that[14]

$$N = M. \qquad (34)$$

That is, under the rotation in momentum space, $(|x\rangle, |y\rangle, |z\rangle)^T$ and $(\hat{\mathbf{k}}_x, \hat{\mathbf{k}}_y, \hat{\mathbf{k}}_z)^T$ (superscript $T$ stands for transpose) transform in the same way.

According to Eqs. (9) and (13a, b), we have

$$\left|S_{\hat{\mathbf{x}}} = +\hbar\right\rangle \equiv \left|S_{\hat{\mathbf{k}}} = +\hbar\right\rangle_{\alpha=0, \beta=\pi/2} = \frac{1}{2}|a\rangle + \frac{1}{\sqrt{2}}|b\rangle + \frac{1}{2}|c\rangle, \qquad (35a)$$

$$\left|S_{\hat{\mathbf{x}}} = -\hbar\right\rangle \equiv \left|S_{\hat{\mathbf{k}}} = -\hbar\right\rangle_{\alpha=0, \beta=\pi/2} = \frac{1}{2}|a\rangle - \frac{1}{\sqrt{2}}|b\rangle + \frac{1}{2}|c\rangle. \qquad (35b)$$

Take the three directions $(\alpha = 0, \beta = \pi/2)$, $(\alpha = \pi/2, \beta = \pi/2)$ and $(\alpha = 0, \beta = 0)$ as the directions of the x, y and z axis of a Cartesian coordinate system respectively. Obviously, the base states $|a\rangle$, $|b\rangle$ and $|c\rangle$ are exactly the commonly used three base states $|S_{\hat{\mathbf{z}}} = +\hbar\rangle$, $|S_{\hat{\mathbf{z}}} = 0\hbar\rangle$ and $|S_{\hat{\mathbf{z}}} = -\hbar\rangle$ of a spin-1 particle, respectively. In quantum mechanics, it is a well-known result that the base states $|x\rangle$, $|y\rangle$ and $|z\rangle$ defined by Eq. (9) transform in the same way as the unit space vectors $\hat{\mathbf{e}}_x$, $\hat{\mathbf{e}}_y$ and $\hat{\mathbf{e}}_z$ under the rotation of space. Here, again, we see the identity of the photon momentum direction and the space direction. This identity allows us to calculate the transformation rules of $|S_{\hat{\mathbf{z}}} = +\hbar\rangle$, $|S_{\hat{\mathbf{z}}} = 0\hbar\rangle$ and $|S_{\hat{\mathbf{z}}} = -\hbar\rangle$ under space rotation by using Eq. (34). [15]

The spin polarization of photons corresponds to the polarization of classical electromagnetic waves. In the above we have obtained the relation between the photon spin plane and the photon momentum direction (Eqs. (23) and (24)). Let us analyze the embodiment of this relation in the classical electromagnetic wave. Usually we talk about the polarization of electromagnetic waves in terms of their electric fields. However, the electromagnetic waves also have magnetic fields. For a uniform planar electromagnetic wave with a given propagation direction, there is a plane determined by the electric field vector and the magnetic field vector corresponding to the propagation direction. This is consistent with the **Assumption Ⅰ**. Suppose we have a superposition of two electromagnetic waves **A** and **B**, both of which are under linear polarization, and have the same amplitude, the same frequency, and the same direction of propagation. The electric field vector of **A** is in the direction of the electromagnetic field vector of **B**, but is advanced or delayed $\pi/2$ in phase compared to the electric field vector of **B**. Correspondingly, an electromagnetic wave with right-handed or left-handed polarization is obtained. The right-handed and left-handed polarizations of the electromagnetic wave correspond to the spin states $|S_{\hat{\mathbf{k}}} = +\hbar\rangle$ and $|S_{\hat{\mathbf{k}}} = -\hbar\rangle$ of photons, respectively. Based on the above analysis, these two states can be written as

$$|S_{\hat{\mathbf{k}}} = +\hbar\rangle = \frac{1}{\sqrt{2}} e^{i\pi/4} (|\hat{\mathbf{E}}\rangle + i|\hat{\mathbf{H}}\rangle), \qquad (36a)$$

$$|S_{\hat{\mathbf{k}}} = -\hbar\rangle = \frac{1}{\sqrt{2}} e^{i\pi/4} (|\hat{\mathbf{E}}\rangle - i|\hat{\mathbf{H}}\rangle). \qquad (36b)$$

where $|\hat{\mathbf{E}}\rangle$ and $|\hat{\mathbf{H}}\rangle$ denote the spin states of the photons corresponding to the electromagnetic waves polarized along the directions of the electric field and of the magnetic field of **A**, respectively. The common phase factor $e^{i\pi/4}$ is introduced so that the direction vectors of the electric and magnetic fields discussed in the following are real numbers. According to Eqs. (36a, b), we have

$$|\hat{\mathbf{E}}\rangle = \frac{1}{\sqrt{2}} e^{-i\pi/4} (|S_{\hat{\mathbf{k}}} = +\hbar\rangle + |S_{\hat{\mathbf{k}}} = -\hbar\rangle), \qquad (37a)$$

$$|\hat{\mathbf{H}}\rangle = \frac{-i}{\sqrt{2}} e^{-i\pi/4} (|S_{\hat{\mathbf{k}}} = +\hbar\rangle - |S_{\hat{\mathbf{k}}} = -\hbar\rangle). \qquad (37b)$$

Substituting them into Eqs. (10a, b), we obtain

$$|\hat{\mathbf{E}}\rangle = \sin\alpha |x\rangle - \cos\alpha |y\rangle, \qquad (38a)$$

$$|\hat{\mathbf{H}}\rangle = \cos\alpha \cos\beta |x\rangle + \sin\alpha \cos\beta |y\rangle - \sin\beta |z\rangle. \qquad (38b)$$

We see above that under the rotation in photon momentum space, $(|x\rangle, |y\rangle, |z\rangle)^T$ and $(\hat{\mathbf{k}}_x, \hat{\mathbf{k}}_y, \hat{\mathbf{k}}_z)^T$ change in the same way. Thus, the following two unit vectors

$$\hat{\mathbf{E}} = \sin\alpha \hat{\mathbf{k}}_x - \cos\alpha \hat{\mathbf{k}}_y, \qquad (39a)$$

$$\hat{\mathbf{H}} = \cos\alpha \cos\beta \hat{\mathbf{k}}_x + \sin\alpha \cos\beta \hat{\mathbf{k}}_y - \sin\beta \hat{\mathbf{k}}_z, \qquad (39b)$$

behave in the same way as the spin states $|\hat{\mathbf{E}}\rangle$ and $|\hat{\mathbf{H}}\rangle$ under space rotation, and thus according the analysis above, they represent the electric and magnetic field directions of the electromagnetic wave **A**. Obviously, the relation between the photon spin planes and the photon momentum directions represented by Eq. (23) means that the relation between the propagation direction of the electromagnetic wave A and its electric and magnetic field directions is

$$\hat{\mathbf{k}} = \hat{\mathbf{E}} \times \hat{\mathbf{H}}. \tag{40}$$

where, the "×" represents the cross product between two vectors. Substituting Eqs. (39a, b) into Eq. (40), we have

$$\hat{\mathbf{k}} = \sin\beta\cos\alpha\,\hat{\mathbf{k}}_\mathbf{x} + \sin\beta\sin\alpha\,\hat{\mathbf{k}}_\mathbf{y} + \cos\beta\,\hat{\mathbf{k}}_\mathbf{z}. \tag{41}$$

It is consistent with Eq. (24). The same result can also be obtained by substituting Eqs. (38a, b) into Eq. (23). The relation between the direction of momentum and the directions of the electric field and the magnetic field for a classical electromagnetic wave, represented by Eq. (40), coincides with that represented by the Poynting's vector (energy flux vector for electromagnetic waves)

$$\vec{\mathbf{S}} = \vec{\mathbf{E}} \times \vec{\mathbf{H}}. \tag{42}$$

Equation (23) is expressed using a special set of base states $|x\rangle$, $|y\rangle$ and $|z\rangle$. The spin plane determined by any two of these three base states is one of the spin planes determined by $|S_{\hat{\mathbf{k}}} = +\hbar\rangle$ and $|S_{\hat{\mathbf{k}}} = -\hbar\rangle$ represented by Eqs. (10a, b) (by selecting appropriate values of α and β). If $|a\rangle$, $|b\rangle$ and $|c\rangle$ in Eqs. (1a, b) are chosen as the base states, then we can prove that both of the spin planes $(|a\rangle,|b\rangle)$ and $(|b\rangle,|c\rangle)$ can not be obtained anyway by choosing α and β values.[16] Thus, $f(|a\rangle,|b\rangle)$ and $f(|b\rangle,|c\rangle)$ do not represent any actual photon momentum direction. For any spin base states $|a'\rangle$, $|b'\rangle$ and $|c'\rangle$, if we use $f(|a'\rangle,|b'\rangle)$, $f(|b'\rangle,|c'\rangle)$ and $f(|c'\rangle,|a'\rangle)$ in the sense that they are expanded as linear combinations of $f(|x\rangle,|y\rangle)$, $f(|y\rangle,|z\rangle)$ and $f(|z\rangle,|x\rangle)$ according to Eq. (23), then we can prove that[17]

$$\begin{aligned}&f(a_{11}|a'\rangle + a_{12}|b'\rangle + a_{13}|c'\rangle, a_{21}|a'\rangle + a_{22}|b'\rangle + a_{23}|c'\rangle)\\&= (a_{11}a_{22} - a_{12}a_{21})f(|a'\rangle,|b'\rangle) + (a_{12}a_{23} - a_{13}a_{22})f(|b'\rangle,|c'\rangle) + (a_{13}a_{21} - a_{11}a_{23})f(|c'\rangle,|a'\rangle).\end{aligned} \tag{43}$$

In summary, in this paper I derived a relation between the momentum directions and the spin planes of photons (Eq. (23)) based on two assumptions on the spin properties of photons. Based on this relation, I discussed the three-dimensionality of the photon momentum, the relation between the photon momentum and its components, and the change law of the spin base states of the spin-1 particles under the rotation in the momentum space. I also discussed the reflection of this relation in classical electromagnetic waves. All the results are consistent with those we are familiar with.

These arguments are to show that the three-dimensionality of the photon momentum is likely to be a property of the photons themselves, and thus the structure of the physical space arises from the structure of the photon momentum space. This view on the space dimension enlightens us that if we observe the physical world by means of another kind of massless particles, such as the gravitons (rest mass 0 and spin 2), we will think that the space is not three-dimensional. For the gravitons, there are 10 independent spin planes in their spin space, so that if the two assumptions about photons are generalized to gravitons in a natural way, then one would conclude that their "momentum" space is 10-dimensional. Considering the enormous difficulties people encounter on quantizing the gravity, the view discussed here provides a new way to think about the problem.

# Supporting Information

## Part A

The U (2) matrix can be expressed generally as

$$C = \begin{pmatrix} a & -b^* \\ b & a^* \end{pmatrix} e^{i\varphi}, \quad (A1)$$

satisfying $aa^* + bb^* = 1$. Therefore, the Eqs. (4a, b) in the main text can be expressed as

$$|1\rangle = C_{11}|S_{\hat{\mathbf{k}}} = +\hbar\rangle + C_{12}|S_{\hat{\mathbf{k}}} = -\hbar\rangle = ae^{i\varphi}|S_{\hat{\mathbf{k}}} = +\hbar\rangle - b^* e^{i\varphi}|S_{\hat{\mathbf{k}}} = -\hbar\rangle, \quad \text{(A2-a)}$$

$$|2\rangle = C_{21}|S_{\hat{\mathbf{k}}} = +\hbar\rangle + C_{22}|S_{\hat{\mathbf{k}}} = -\hbar\rangle = be^{i\varphi}|S_{\hat{\mathbf{k}}} = +\hbar\rangle + a^* e^{i\varphi}|S_{\hat{\mathbf{k}}} = -\hbar\rangle. \quad \text{(A2-b)}$$

Define two new base states:

$$|S'_{\hat{\mathbf{k}}} = +\hbar\rangle = e^{i\varphi}|S_{\hat{\mathbf{k}}} = +\hbar\rangle, \quad \text{(A3-a)}$$

$$|S'_{\hat{\mathbf{k}}} = -\hbar\rangle = e^{i\varphi}|S_{\hat{\mathbf{k}}} = -\hbar\rangle. \quad \text{(A3-b)}$$

We have

$$|1\rangle = a|S'_{\hat{\mathbf{k}}} = +\hbar\rangle - b^*|S'_{\hat{\mathbf{k}}} = -\hbar\rangle, \quad \text{(A4-a)}$$

$$|2\rangle = b|S'_{\hat{\mathbf{k}}} = +\hbar\rangle + a^*|S'_{\hat{\mathbf{k}}} = -\hbar\rangle. \quad \text{(A4-b)}$$

This is equivalent to the requirement $\det C = 1$.

Similarly, if we define two new base states:

$$|S''_{\hat{\mathbf{k}}} = +\hbar\rangle = e^{i(\varphi-\pi/2)}|S_{\hat{\mathbf{k}}} = +\hbar\rangle, \quad \text{(A5-a)}$$

$$|S''_{\hat{\mathbf{k}}} = -\hbar\rangle = e^{i(\varphi-\pi/2)}|S_{\hat{\mathbf{k}}} = -\hbar\rangle. \quad \text{(A5-b)}$$

then we have

$$|1\rangle = ai|S''_{\hat{\mathbf{k}}} = +\hbar\rangle - b^* i|S''_{\hat{\mathbf{k}}} = -\hbar\rangle, \quad \text{(A6-a)}$$

$$|2\rangle = bi|S''_{\hat{\mathbf{k}}} = +\hbar\rangle + a^* i|S''_{\hat{\mathbf{k}}} = -\hbar\rangle. \quad \text{(A6-b)}$$

This is equivalent to the requirement $\det C = -1$.

## Part B

**Proofs for Eqs. (12b, c) in the main text.**

$$f(|S_{\hat{x}}=+\hbar\rangle,|S_{\hat{x}}=-\hbar\rangle) = f(-\frac{i}{\sqrt{2}}e^{-i\pi/4}|y\rangle+\frac{1}{\sqrt{2}}e^{-i\pi/4}|z\rangle, -\frac{i}{\sqrt{2}}e^{-i\pi/4}|y\rangle-\frac{1}{\sqrt{2}}e^{-i\pi/4}|z\rangle)$$

$$=[(-\frac{i}{\sqrt{2}}e^{-i\pi/4})(-\frac{1}{\sqrt{2}}e^{-i\pi/4})-\frac{1}{\sqrt{2}}e^{-i\pi/4}(-\frac{i}{\sqrt{2}}e^{-i\pi/4})]f(|y\rangle,|z\rangle) \quad (B1)$$

$$=f(|y\rangle,|z\rangle).$$

$$f(|S_{\hat{y}}=+\hbar\rangle,|S_{\hat{y}}=-\hbar\rangle) = f(\frac{i}{\sqrt{2}}e^{-i\pi/4}|x\rangle+\frac{1}{\sqrt{2}}e^{-i\pi/4}|z\rangle, \frac{i}{\sqrt{2}}e^{-i\pi/4}|x\rangle-\frac{1}{\sqrt{2}}e^{-i\pi/4}|z\rangle)$$

$$=[\frac{1}{\sqrt{2}}e^{-i\pi/4}\frac{i}{\sqrt{2}}e^{-i\pi/4}-\frac{i}{\sqrt{2}}e^{-i\pi/4}(-\frac{1}{\sqrt{2}}e^{-i\pi/4})]f(|z\rangle,|x\rangle) \quad (B2)$$

$$= f(|z\rangle,|x\rangle).$$

## Part C

According to the Eqs. (10a, b) and (16a, b) in the main text, we have

$$d_{11}=\frac{1}{\sqrt{2}}(i\sin\alpha-\cos\alpha\cos\beta)e^{-i\pi/4}, d_{12}=\frac{1}{\sqrt{2}}(-i\cos\alpha-\sin\alpha\cos\beta)e^{-i\pi/4}, d_{13}=\frac{1}{\sqrt{2}}\sin\beta e^{-i\pi/4},$$

$$d_{21}=\frac{1}{\sqrt{2}}(i\sin\alpha+\cos\alpha\cos\beta)e^{-i\pi/4}, d_{22}=\frac{1}{\sqrt{2}}(-i\cos\alpha+\sin\alpha\cos\beta)e^{-i\pi/4}, d_{23}=-\frac{1}{\sqrt{2}}\sin\beta e^{-i\pi/4}.$$

(C1)

The condition $d_{11}=d_{21}=0$ is that

$$\sin\alpha=0 \quad \Rightarrow \alpha=0 \text{ or } \pi, \quad (C2)$$

$$\cos\alpha\cos\beta=0 \quad \Rightarrow \beta=\pi/2. \quad (C3)$$

The condition $d_{12}=d_{22}=0$ is that

$$\cos\alpha=0 \quad \Rightarrow \alpha=\pi/2 \text{ or } 3\pi/2, \quad (C4)$$

$$\sin\alpha\cos\beta=0 \quad \Rightarrow \beta=\pi/2. \quad (C5)$$

The condition $d_{13}=d_{23}=0$ is that

$$\sin\beta=0 \quad \Rightarrow \beta=0 \text{ or } \pi. \quad (C6)$$

# Part D

**Proposition**:

The correspondence given by the following equation from the spin planes of a photon to the photon momentum directions is one-to-one:

$$\hat{\mathbf{k}} = f(\frac{1}{\sqrt{2}}(i\sin\alpha - \cos\alpha\cos\beta)e^{-i\pi/4}|x\rangle + \frac{1}{\sqrt{2}}(-i\cos\alpha - \sin\alpha\cos\beta)e^{-i\pi/4}|y\rangle + \frac{1}{\sqrt{2}}\sin\beta e^{-i\pi/4}|z\rangle,$$

$$\frac{1}{\sqrt{2}}(i\sin\alpha + \cos\alpha\cos\beta)e^{-i\pi/4}|x\rangle + \frac{1}{\sqrt{2}}(-i\cos\alpha + \sin\alpha\cos\beta)e^{-i\pi/4}|y\rangle - \frac{1}{\sqrt{2}}\sin\beta e^{-i\pi/4}|z\rangle)$$

$$= \sin\beta\cos\alpha f(|y\rangle,|z\rangle) + \sin\beta\sin\alpha f(|z\rangle,|x\rangle) + \cos\beta f(|x\rangle,|y\rangle)$$

$$= \sin\beta\cos\alpha \hat{\mathbf{k}}_\mathbf{x} + \sin\beta\sin\alpha \hat{\mathbf{k}}_\mathbf{y} + \cos\beta \hat{\mathbf{k}}_\mathbf{z}.$$

(D1)

**Proof**:

If the two sets of base states $(|S_{\hat{\mathbf{k}}_1} = +\hbar\rangle, |S_{\hat{\mathbf{k}}_1} = \hbar\rangle)$ and $(|S_{\hat{\mathbf{k}}_2} = +\hbar\rangle, |S_{\hat{\mathbf{k}}_2} = \hbar\rangle)$ determine a same spin planes, thus, according to the analysis in the main text, we have

$$\begin{pmatrix} |S_{\hat{\mathbf{k}}_2} = +\hbar\rangle \\ |S_{\hat{\mathbf{k}}_2} = -\hbar\rangle \end{pmatrix} = \begin{pmatrix} C_{11} & C_{12} \\ C_{21} & C_{22} \end{pmatrix} \begin{pmatrix} |S_{\hat{\mathbf{k}}_1} = +\hbar\rangle \\ |S_{\hat{\mathbf{k}}_1} = -\hbar\rangle \end{pmatrix} = \begin{pmatrix} C_{11}|S_{\hat{\mathbf{k}}_1} = +\hbar\rangle + C_{12}|S_{\hat{\mathbf{k}}_1} = -\hbar\rangle \\ C_{21}|S_{\hat{\mathbf{k}}_1} = +\hbar\rangle + C_{22}|S_{\hat{\mathbf{k}}_1} = -\hbar\rangle \end{pmatrix},$$  (D2)

where, $C_{i,j}$ $(i, j = 1, 2)$ are the elements of a $SU(2)$ matrix $C$. According to the Eq. (8) in the main text and $\det C = 1$, we have

$$f(|S_{\hat{\mathbf{k}}_2} = +\hbar\rangle, |S_{\hat{\mathbf{k}}_2} = \hbar\rangle)$$

$$= f(C_{11}|S_{\hat{\mathbf{k}}_1} = +\hbar\rangle + C_{12}|S_{\hat{\mathbf{k}}_1} = -\hbar\rangle, C_{21}|S_{\hat{\mathbf{k}}_1} = +\hbar\rangle + C_{22}|S_{\hat{\mathbf{k}}_1} = -\hbar\rangle)$$

$$= (C_{11}C_{22} - C_{12}C_{21}) f(|S_{\hat{\mathbf{k}}_1} = +\hbar\rangle, |S_{\hat{\mathbf{k}}_1} = -\hbar\rangle)$$

$$= f(|S_{\hat{\mathbf{k}}_1} = +\hbar\rangle, |S_{\hat{\mathbf{k}}_1} = -\hbar\rangle).$$

(D3)

Therefore, the same spin plane corresponds to the same direction of momentum.

Below we prove that the same direction of momentum corresponds to the same spin plane. Suppose $(\alpha_1, \beta_1)$ and $(\alpha_2, \beta_2)$ represent a same direction of momentum, and thus according to Eq. (D1) we have

$$\sin\beta_1 \cos\alpha_1 = \sin\beta_2 \cos\alpha_2, \tag{D4-a}$$

$$\sin\beta_1 \sin\alpha_1 = \sin\beta_2 \sin\alpha_2, \tag{D4-b}$$

$$\cos\beta_1 = \cos\beta_2. \tag{D4-c}$$

We need to prove that $(|S_{\hat{\mathbf{k}}_1} = +\hbar\rangle_{\alpha=\alpha_1,\beta=\beta_1}, |S_{\hat{\mathbf{k}}_1} = \hbar\rangle_{\alpha=\alpha_1,\beta=\beta_1})$ and

$(\left|S_{\hat{\mathbf{k}}_2} = +\hbar\right\rangle_{\alpha=\alpha_2, \beta=\beta_2}, \left|S_{\hat{\mathbf{k}}_2} = -\hbar\right\rangle_{\alpha=\alpha_2, \beta=\beta_2})$ are a same spin plane.

According to Eq. (D4-c) and $\beta_1, \beta_2 \in [0, \pi]$, we have

$$\beta_1 = \beta_2. \tag{D5}$$

1) If $\beta_1 = \beta_2 \neq 0$ or $\pi$, according to Eqs. (D4-a) and (D4-b), we have

$$\cos\alpha_1 = \cos\alpha_2, \tag{D6}$$

$$\sin\alpha_1 = \sin\alpha_2. \tag{D7}$$

And because $\alpha_1, \alpha_2 \in [0, 2\pi)$, we have

$$\alpha_1 = \alpha_2. \tag{D8}$$

According to Eqs. (D5) and (D8), obviously, $(\left|S_{\hat{\mathbf{k}}_1} = +\hbar\right\rangle_{\alpha=\alpha_1, \beta=\beta_1}, \left|S_{\hat{\mathbf{k}}_1} = -\hbar\right\rangle_{\alpha=\alpha_1, \beta=\beta_1})$ and $(\left|S_{\hat{\mathbf{k}}_2} = +\hbar\right\rangle_{\alpha=\alpha_2, \beta=\beta_2}, \left|S_{\hat{\mathbf{k}}_2} = -\hbar\right\rangle_{\alpha=\alpha_2, \beta=\beta_2})$ represent a same spin plane.

2) If $\beta_1 = \beta_2 = 0$, $\alpha_1, \alpha_2$ can take any values in the range of $[0, 2\pi)$. According to Eqs. (10a, b) in the main text, we have

$$\left|S_{\hat{\mathbf{k}}_1} = +\hbar\right\rangle_{\alpha=\alpha_1, \beta=0} = \frac{1}{\sqrt{2}}(i\sin\alpha_1 - \cos\alpha_1)e^{-i\pi/4}|x\rangle + \frac{1}{\sqrt{2}}(-i\cos\alpha_1 - \sin\alpha_1)e^{-i\pi/4}|y\rangle, \tag{D9-a}$$

$$\left|S_{\hat{\mathbf{k}}_1} = -\hbar\right\rangle_{\alpha=\alpha_1, \beta=0} = \frac{1}{\sqrt{2}}(i\sin\alpha_1 + \cos\alpha_1)e^{-i\pi/4}|x\rangle + \frac{1}{\sqrt{2}}(-i\cos\alpha_1 + \sin\alpha_1)e^{-i\pi/4}|y\rangle, \tag{D9-b}$$

$$\left|S_{\hat{\mathbf{k}}_2} = +\hbar\right\rangle_{\alpha=\alpha_2, \beta=0} = \frac{1}{\sqrt{2}}(i\sin\alpha_2 - \cos\alpha_2)e^{-i\pi/4}|x\rangle + \frac{1}{\sqrt{2}}(-i\cos\alpha_2 - \sin\alpha_2)e^{-i\pi/4}|y\rangle, \tag{D9-c}$$

$$\left|S_{\hat{\mathbf{k}}_2} = -\hbar\right\rangle_{\alpha=\alpha_2, \beta=0} = \frac{1}{\sqrt{2}}(i\sin\alpha_2 + \cos\alpha_2)e^{-i\pi/4}|x\rangle + \frac{1}{\sqrt{2}}(-i\cos\alpha_2 + \sin\alpha_2)e^{-i\pi/4}|y\rangle. \tag{D9-d}$$

Now we show that there is a $SU(2)$ matrix $C'$, satisfying

$$\begin{pmatrix} \left|S_{\hat{\mathbf{k}}_2} = +\hbar\right\rangle_{\alpha=\alpha_2, \beta=0} \\ \left|S_{\hat{\mathbf{k}}_2} = -\hbar\right\rangle_{\alpha=\alpha_2, \beta=0} \end{pmatrix} = C' \begin{pmatrix} \left|S_{\hat{\mathbf{k}}_1} = +\hbar\right\rangle_{\alpha=\alpha_1, \beta=0} \\ \left|S_{\hat{\mathbf{k}}_1} = -\hbar\right\rangle_{\alpha=\alpha_1, \beta=0} \end{pmatrix} = \begin{pmatrix} C'_{11} & C'_{12} \\ C'_{21} & C'_{22} \end{pmatrix} \begin{pmatrix} \left|S_{\hat{\mathbf{k}}_1} = +\hbar\right\rangle_{\alpha=\alpha_1, \beta=0} \\ \left|S_{\hat{\mathbf{k}}_1} = -\hbar\right\rangle_{\alpha=\alpha_1, \beta=0} \end{pmatrix}. \tag{D10}$$

That is

$$\frac{1}{\sqrt{2}}(i\sin\alpha_2 - \cos\alpha_2)e^{-i\pi/4}|x\rangle + \frac{1}{\sqrt{2}}(-i\cos\alpha_2 - \sin\alpha_2)e^{-i\pi/4}|y\rangle$$

$$= \frac{1}{\sqrt{2}}[C'_{11}(i\sin\alpha_1 - \cos\alpha_1) + C'_{12}(i\sin\alpha_1 + \cos\alpha_1)]e^{-i\pi/4}|x\rangle \tag{D11-a}$$

$$+ \frac{1}{\sqrt{2}}[C'_{11}(-i\cos\alpha_1 - \sin\alpha_1) + C'_{12}(-i\cos\alpha_1 + \sin\alpha_1)]e^{-i\pi/4}|y\rangle,$$

$$\frac{1}{\sqrt{2}}(i\sin\alpha_2 + \cos\alpha_2)e^{-i\pi/4}|x\rangle + \frac{1}{\sqrt{2}}(-i\cos\alpha_2 + \sin\alpha_2)e^{-i\pi/4}|y\rangle$$

$$= \frac{1}{\sqrt{2}}[C'_{21}(i\sin\alpha_1 - \cos\alpha_1) + C'_{22}(i\sin\alpha_1 + \cos\alpha_1)]e^{-i\pi/4}|x\rangle \quad \text{(D11-b)}$$

$$+ \frac{1}{\sqrt{2}}[C'_{21}(-i\cos\alpha_1 - \sin\alpha_1) + C'_{22}(-i\cos\alpha_1 + \sin\alpha_1)]e^{-i\pi/4}|y\rangle.$$

Thus we have

$$i\sin\alpha_2 - \cos\alpha_2 = i\sin\alpha_1(C'_{11} + C'_{12}) + \cos\alpha_1(C'_{12} - C'_{11}), \quad \text{(D12-a)}$$

$$-i\cos\alpha_2 - \sin\alpha_2 = -i\cos\alpha_1(C'_{11} + C'_{12}) + \sin\alpha_1(C'_{12} - C'_{11}), \quad \text{(D12-b)}$$

$$i\sin\alpha_2 + \cos\alpha_2 = i\sin\alpha_1(C'_{21} + C'_{22}) + \cos\alpha_1(C'_{22} - C'_{21}), \quad \text{(D12-c)}$$

$$i\cos\alpha_2 + \sin\alpha_2 = -i\cos\alpha_1(C'_{11} + C'_{12}) + \sin\alpha_1(C'_{12} - C'_{11}). \quad \text{(D12-d)}$$

The solutions of Eqs. (D12-a, b, c, d) are

$$C'_{11} = \sin\alpha_1\sin\alpha_2 + \cos\alpha_1\cos\alpha_2 + i(\sin\alpha_1\cos\alpha_2 - \cos\alpha_1\sin\alpha_2), \quad \text{(D13-a)}$$

$$C'_{12} = 0, \quad \text{(D13-b)}$$

$$C'_{21} = 0, \quad \text{(D13-c)}$$

$$C'_{22} = \sin\alpha_1\sin\alpha_2 + \cos\alpha_1\cos\alpha_2 + i(\cos\alpha_1\sin\alpha_2 - \sin\alpha_1\cos\alpha_2), \quad \text{(D13-d)}$$

namely,

$$C' = \begin{pmatrix} \sin\alpha_1\sin\alpha_2 + \cos\alpha_1\cos\alpha_2 + i(\sin\alpha_1\cos\alpha_2 - \cos\alpha_1\sin\alpha_2) & 0 \\ 0 & \sin\alpha_1\sin\alpha_2 + \cos\alpha_1\cos\alpha_2 + i(\cos\alpha_1\sin\alpha_2 - \sin\alpha_1\cos\alpha_2) \end{pmatrix}.$$

(D14)

It is easy to verify that, $\det C' = 1$ and $C'C'^\dagger = 1$. Namely, $C'$ is a $SU(2)$ matrix. Therefore,

$$\left(\left|S_{\hat{\mathbf{k}}_1} = +\hbar\right\rangle_{\alpha=\alpha_1,\beta=0}, \left|S_{\hat{\mathbf{k}}_1} = \hbar\right\rangle_{\alpha=\alpha_1,\beta=0}\right) \quad \text{and} \quad \left(\left|S_{\hat{\mathbf{k}}_2} = +\hbar\right\rangle_{\alpha=\alpha_2,\beta=0}, \left|S_{\hat{\mathbf{k}}_2} = \hbar\right\rangle_{\alpha=\alpha_2,\beta=0}\right)$$

represent a same spin plane.

3) If $\beta_1 = \beta_2 = \pi$, $\alpha_1,\alpha_2$ can take any values in the range of $[0,2\pi)$. Similarly, we can prove that the $SU(2)$ matrix

$$C'' = \begin{pmatrix} \sin\alpha_1\sin\alpha_2 + \cos\alpha_1\cos\alpha_2 + i(\cos\alpha_1\sin\alpha_2 - \sin\alpha_1\cos\alpha_2) & 0 \\ 0 & \sin\alpha_1\sin\alpha_2 + \cos\alpha_1\cos\alpha_2 + i(\sin\alpha_1\sin\alpha_2 - \cos\alpha_1\sin\alpha_2) \end{pmatrix}$$

(D15)

satisfying

$$\begin{pmatrix} \left|S_{\hat{\mathbf{k}}_2} = +\hbar\right\rangle_{\alpha=\alpha_2,\beta=\pi} \\ \left|S_{\hat{\mathbf{k}}_2} = -\hbar\right\rangle_{\alpha=\alpha_2,\beta=\pi} \end{pmatrix} = \begin{pmatrix} C''_{11} & C''_{12} \\ C''_{21} & C''_{22} \end{pmatrix} \begin{pmatrix} \left|S_{\hat{\mathbf{k}}_1} = +\hbar\right\rangle_{\alpha=\alpha_1,\beta=\pi} \\ \left|S_{\hat{\mathbf{k}}_1} = -\hbar\right\rangle_{\alpha=\alpha_1,\beta=\pi} \end{pmatrix}. \quad \text{(D16)}$$

Therefore, $(\left|S_{\hat{\mathbf{k}}_1} = +\hbar\right\rangle_{\alpha=\alpha_1,\beta=\pi}, \left|S_{\hat{\mathbf{k}}_1} = \hbar\right\rangle_{\alpha=\alpha_1,\beta=\pi})$ and $(\left|S_{\hat{\mathbf{k}}_2} = +\hbar\right\rangle_{\alpha=\alpha_2,\beta=\pi}, \left|S_{\hat{\mathbf{k}}_2} = \hbar\right\rangle_{\alpha=\alpha_2,\beta=\pi})$ represent a same spin plane.

Therefore, the same direction of momentum corresponds to the same spin plane.

In summary, the correspondence from the spin planes of a photon to the directions of photon momentum given by the Eq. (D1) in the main text is one-to-one.

## Part E

**Proposition:**

It is impossible to decompose all the photon momentums into two selected directions.

**Proof:**

Suppose it is possible to decompose all the photon momentums into two selected directions, and thus we have

$$\begin{aligned}
&f(\left|S_{\hat{\mathbf{k}}} = +\hbar\right\rangle, \left|S_{\hat{\mathbf{k}}} = -\hbar\right\rangle) \\
&= g_1''(\alpha,\beta) f(\left|S_{\hat{\mathbf{k}}} = +\hbar\right\rangle_{\alpha=\alpha_1',\beta=\beta_1'}, \left|S_{\hat{\mathbf{k}}} = -\hbar\right\rangle_{\alpha=\alpha_1',\beta=\beta_1'}) \\
&+ g_1''(\alpha,\beta) f(\left|S_{\hat{\mathbf{k}}} = +\hbar\right\rangle_{\alpha=\alpha_2',\beta=\beta_2'}, \left|S_{\hat{\mathbf{k}}} = -\hbar\right\rangle_{\alpha=\alpha_2',\beta=\beta_2'}).
\end{aligned} \quad \text{(E1)}$$

Expand the both sides of the Eq. (E1) using the Eq. (24) in the main text, according to the one-to-one correspondence proved in Part D, we have

$$\sin\beta\cos\alpha = g_1''(\alpha,\beta)\sin\beta_1'\cos\alpha_1' + g_2''(\alpha,\beta)\sin\beta_2'\cos\alpha_2', \quad \text{(E2-a)}$$

$$\sin\beta\sin\alpha = g_1''(\alpha,\beta)\sin\beta_1'\sin\alpha_1' + g_2''(\alpha,\beta)\sin\beta_2'\sin\alpha_2', \quad \text{(E2-b)}$$

$$\cos\beta = g_1''(\alpha,\beta)\cos\beta_1' + g_2''(\alpha,\beta)\cos\beta_2'. \quad \text{(E2-c)}$$

Given $\alpha_1'$, $\beta_1'$, $\alpha_2'$, $\beta_2'$, the Eqs. (E2-a, b and c) are equations about $g_1''(\alpha,\beta)$ and $g_2''(\alpha,\beta)$. Because $\sin\beta\cos\alpha$, $\sin\beta\sin\alpha$ and $\cos\beta$ are three linearly independent functions, these equations have no solution. We can also prove this by solving $g_1''(\alpha,\beta)$ and $g_2''(\alpha,\beta)$ from Eqs. (E2-a, b) and substituting them into Eq. (E2-c).

Therefore, it is impossible to decompose all the photon momentums into two selected directions.

## Part F

The expressions of the matrices $B_1$, $B_2$ and $B_3$:

$$B_1 = \begin{pmatrix} \sin\beta\cos\alpha & \sin\beta_2\cos\alpha_2 & \sin\beta_3\cos\alpha_3 \\ \sin\beta\sin\alpha & \sin\beta_2\sin\alpha_2 & \sin\beta_3\sin\alpha_3 \\ \cos\beta & \cos\beta_2 & \cos\beta_3 \end{pmatrix}, \tag{F1}$$

$$B_2 = \begin{pmatrix} \sin\beta_1\cos\alpha_1 & \sin\beta\cos\alpha & \sin\beta_3\cos\alpha_3 \\ \sin\beta_1\sin\alpha_1 & \sin\beta\sin\alpha & \sin\beta_3\sin\alpha_3 \\ \cos\beta_1 & \cos\beta & \cos\beta_3 \end{pmatrix}, \tag{F2}$$

$$B_3 = \begin{pmatrix} \sin\beta_1\cos\alpha_1 & \sin\beta_2\cos\alpha_2 & \sin\beta\cos\alpha \\ \sin\beta_1\sin\alpha_1 & \sin\beta_2\sin\alpha_2 & \sin\beta\sin\alpha \\ \cos\beta_1 & \cos\beta_2 & \cos\beta \end{pmatrix}. \tag{F3}$$

## Part G

**Proposition:**

The condition $\det B \neq 0$ exactly requires that $\hat{\mathbf{k}}_1$, $\hat{\mathbf{k}}_2$ and $\hat{\mathbf{k}}_3$ are not coplanar.

**Proof:**

According to the Eq. (28) in the main text, we have

$$\begin{aligned} \det B = &\cos\beta_1 \sin\beta_2 \sin\beta_3 (\cos\alpha_2 \sin\alpha_3 - \cos\alpha_3 \sin\alpha_2) \\ &- \sin\beta_1 \cos\beta_2 \sin\beta_3 (\sin\alpha_3 \cos\alpha_1 - \sin\alpha_1 \cos\alpha_3) \\ &+ \sin\beta_1 \sin\beta_2 \cos\beta_3 (\cos\alpha_1 \sin\alpha_2 - \cos\alpha_2 \sin\alpha_1). \end{aligned} \tag{G1}$$

If $\hat{\mathbf{k}}_1$, $\hat{\mathbf{k}}_2$ and $\hat{\mathbf{k}}_3$ are coplanar, we have

$$(\hat{\mathbf{k}}_1 \times \hat{\mathbf{k}}_2) \cdot \hat{\mathbf{k}}_3 = 0. \tag{G2}$$

Thus we have

$$\begin{aligned} &(\sin\beta_1 \cos\alpha_1 \hat{\mathbf{k}}_x + \sin\beta_1 \sin\alpha_1 \hat{\mathbf{k}}_y + \cos\beta_1 \hat{\mathbf{k}}_z) \\ &\times (\sin\beta_2 \cos\alpha_2 \hat{\mathbf{k}}_x + \sin\beta_2 \sin\alpha_2 \hat{\mathbf{k}}_y + \cos\beta_2 \hat{\mathbf{k}}_z) \\ &\cdot (\sin\beta_3 \cos\alpha_3 \hat{\mathbf{k}}_x + \sin\beta_3 \sin\alpha_3 \hat{\mathbf{k}}_y + \cos\beta_3 \hat{\mathbf{k}}_z) \\ &= \cos\beta_1 \sin\beta_2 \sin\beta_3 (\cos\alpha_2 \sin\alpha_3 - \cos\alpha_3 \sin\alpha_2) \\ &- \sin\beta_1 \cos\beta_2 \sin\beta_3 (\sin\alpha_3 \cos\alpha_1 - \sin\alpha_1 \cos\alpha_3) \\ &+ \sin\beta_1 \sin\beta_2 \cos\beta_3 (\cos\alpha_1 \sin\alpha_2 - \cos\alpha_2 \sin\alpha_1) = 0. \end{aligned} \tag{G3}$$

Obviously, Eq. (G1) is just Eq. (G3). Therefore, the condition $\det B \neq 0$ exactly requires that $\hat{\mathbf{k}}_1$, $\hat{\mathbf{k}}_2$ and $\hat{\mathbf{k}}_3$ are not coplanar.

# Part H

**Proposition:**

If $\hat{\mathbf{k}}_1$, $\hat{\mathbf{k}}_2$ and $\hat{\mathbf{k}}_3$ are three mutually perpendicular directions then the equation

$$\sum_{i=1}^{3} g_i'^{2}(\alpha,\beta) = 1$$

holds, and vice versa.

**Proof:**

The Eqs. (27a, b and c) in the main text can be written as

$$\begin{pmatrix} \sin\beta\cos\alpha \\ \sin\beta\sin\alpha \\ \cos\beta \end{pmatrix} = B \begin{pmatrix} g_1'(\alpha,\beta) \\ g_2'(\alpha,\beta) \\ g_3'(\alpha,\beta) \end{pmatrix}, \tag{H1}$$

where,

$$B = \begin{pmatrix} \sin\beta_1\cos\alpha_1 & \sin\beta_2\cos\alpha_2 & \sin\beta_3\cos\alpha_3 \\ \sin\beta_1\sin\alpha_1 & \sin\beta_2\sin\alpha_2 & \sin\beta_3\sin\alpha_3 \\ \cos\beta_1 & \cos\beta_2 & \cos\beta_3 \end{pmatrix}. \tag{H2}$$

Multiply both sides of Eq. (H1) from left by the transpose of Eq. (H1), and we have

$$\begin{aligned}&(\sin\beta\cos\alpha \quad \sin\beta\sin\alpha \quad \cos\beta)\begin{pmatrix} \sin\beta\cos\alpha \\ \sin\beta\sin\alpha \\ \cos\beta \end{pmatrix} \\ &= (g_1'(\alpha,\beta) \quad g_2'(\alpha,\beta) \quad g_3'(\alpha,\beta))B^T B \begin{pmatrix} g_1'(\alpha,\beta) \\ g_2'(\alpha,\beta) \\ g_3'(\alpha,\beta) \end{pmatrix}.\end{aligned} \tag{H3}$$

Because

$$(\sin\beta\cos\alpha)^2 + (\sin\beta\sin\alpha)^2 + (\cos\beta)^2 = 1, \tag{H4}$$

we have

$$(g_1'(\alpha,\beta) \quad g_2'(\alpha,\beta) \quad g_3'(\alpha,\beta))B^T B \begin{pmatrix} g_1'(\alpha,\beta) \\ g_2'(\alpha,\beta) \\ g_3'(\alpha,\beta) \end{pmatrix} = 1. \tag{H5}$$

According to Eq. (H2), we have

$$B^T B$$

$$= \begin{pmatrix} \sin\beta_1\cos\alpha_1 & \sin\beta_1\sin\alpha_1 & \cos\beta_1 \\ \sin\beta_2\cos\alpha_2 & \sin\beta_2\sin\alpha_2 & \cos\beta_2 \\ \sin\beta_3\cos\alpha_3 & \sin\beta_3\sin\alpha_3 & \cos\beta_3 \end{pmatrix} \begin{pmatrix} \sin\beta_1\cos\alpha_1 & \sin\beta_2\cos\alpha_2 & \sin\beta_3\cos\alpha_3 \\ \sin\beta_1\sin\alpha_1 & \sin\beta_2\sin\alpha_2 & \sin\beta_3\sin\alpha_3 \\ \cos\beta_1 & \cos\beta_2 & \cos\beta_3 \end{pmatrix}$$

$$= \begin{pmatrix} 1 & (B^T B)_{12} & (B^T B)_{13} \\ (B^T B)_{12} & 1 & (B^T B)_{23} \\ (B^T B)_{13} & (B^T B)_{23} & 1 \end{pmatrix},$$

(H6)

where,

$$(B^T B)_{12} = \sin\beta_1\sin\beta_2\cos\alpha_1\cos\alpha_2 + \sin\beta_1\sin\beta_2\sin\alpha_1\sin\alpha_2 + \cos\beta_1\cos\beta_2, \quad \text{(H7-a)}$$

$$(B^T B)_{13} = \sin\beta_1\sin\beta_3\cos\alpha_1\cos\alpha_3 + \sin\beta_1\sin\beta_3\sin\alpha_1\sin\alpha_3 + \cos\beta_1\cos\beta_3, \quad \text{(H7-b)}$$

$$(B^T B)_{23} = \sin\beta_2\sin\beta_3\cos\alpha_2\cos\alpha_3 + \sin\beta_2\sin\beta_3\sin\alpha_2\sin\alpha_3 + \cos\beta_2\cos\beta_3. \quad \text{(H7-c)}$$

If $\hat{\mathbf{k}}_i(\alpha_i, \beta_i)$ and $\hat{\mathbf{k}}_j(\alpha_j, \beta_j)$ $(i, j = 1, 2, 3)$ are two mutually perpendicular directions, then we have

$$\hat{\mathbf{k}}_i(\alpha_i, \beta_i) \cdot \hat{\mathbf{k}}_j(\alpha_j, \beta_j) = 0, \quad \text{(H8)}$$

and thus

$$(\sin\beta_i\cos\alpha_i\hat{\mathbf{k}}_x + \sin\beta_i\sin\alpha_i\hat{\mathbf{k}}_y + \cos\beta_i\hat{\mathbf{k}}_z) \cdot (\sin\beta_j\cos\alpha_j\hat{\mathbf{k}}_x + \sin\beta_j\sin\alpha_j\hat{\mathbf{k}}_y + \cos\beta_j\hat{\mathbf{k}}_z)$$
$$= \sin\beta_i\sin\beta_j\cos\alpha_i\cos\alpha_j + \sin\beta_i\sin\beta_j\sin\alpha_i\sin\alpha_j + \cos\beta_i\cos\beta_j = 0$$

(H9)

Obviously, if $\hat{\mathbf{k}}_1$, $\hat{\mathbf{k}}_2$ and $\hat{\mathbf{k}}_3$ are three mutually perpendicular directions, according to Eqs. (H9) and (H7-a, b and c), we have

$$(B^T B)_{12} = (B^T B)_{13} = (B^T B)_{23} = 0, \quad \text{(H10)}$$

and thus

$$B^T B = I. \quad \text{(H11)}$$

Substitute Eq. (H11) into Eq. (H5), and we have

$$g_1'(\alpha, \beta)^2 + g_2'(\alpha, \beta)^2 + g_3'(\alpha, \beta)^2 = 1. \quad \text{(H12)}$$

Therefore, if $\hat{\mathbf{k}}_1$, $\hat{\mathbf{k}}_2$ and $\hat{\mathbf{k}}_3$ are three mutually perpendicular directions, we have

$$\sum_{i=1}^{3} g_i'^2(\alpha, \beta) = 1.$$

Below we prove that, if $\sum_{i=1}^{3} g_i'^2 = 1$, $\hat{\mathbf{k}}_1$, $\hat{\mathbf{k}}_2$ and $\hat{\mathbf{k}}_3$ are three mutually perpendicular directions.

Because $\det B \neq 0$ (otherwise Eq. (H1) has no solution), the inverse of the matrix B, $B^{-1}$, always exists. According to Eq. (H1), we have

$$B^{-1}\begin{pmatrix} \sin\beta\cos\alpha \\ \sin\beta\sin\alpha \\ \cos\beta \end{pmatrix} = \begin{pmatrix} g_1'(\alpha,\beta) \\ g_2'(\alpha,\beta) \\ g_3'(\alpha,\beta) \end{pmatrix}. \tag{H13}$$

The transpose of Eq. (H13) is

$$(\sin\beta\cos\alpha \quad \sin\beta\sin\alpha \quad \cos\beta)(B^{-1})^T = (g_1'(\alpha,\beta) \quad g_2'(\alpha,\beta) \quad g_3'(\alpha,\beta)). \tag{H14}$$

Multiplying the Eq. (H13) from left by Eq. (14), and utilizing $\sum_{i=1}^{3} g_i'^2 = 1$, we have

$$(\sin\beta\cos\alpha \quad \sin\beta\sin\alpha \quad \cos\beta)(B^{-1})^T B^{-1} \begin{pmatrix} \sin\beta\cos\alpha \\ \sin\beta\sin\alpha \\ \cos\beta \end{pmatrix} = 1. \tag{H15}$$

Let $D = (B^{-1})^T B^{-1}$. The Eq. (H15) holds for all possible values of $\alpha, \beta$. Take $\alpha = 0, \beta = 0$, we have

$$(0 \quad 0 \quad 1)\begin{pmatrix} D_{11} & D_{12} & D_{13} \\ D_{21} & D_{22} & D_{23} \\ D_{31} & D_{32} & D_{33} \end{pmatrix}\begin{pmatrix} 0 \\ 0 \\ 1 \end{pmatrix} = D_{33} = 1. \tag{H16}$$

Similarly, taking $\alpha = 0, \beta = \pi/2$ and $\alpha = \pi/2, \beta = \pi/2$ respectively, we have

$$D_{11} = D_{22} = 1. \tag{H17}$$

Because $D^T = (B^{-1})^T B^{-1} = D$, $D$ is a symmetric matrix. According to Eqs. (H16) and (H17), the Eq. (15) can be written as

$$(\sin\beta\cos\alpha \quad \sin\beta\sin\alpha \quad \cos\beta)\begin{pmatrix} 1 & D_{12} & D_{13} \\ D_{12} & 1 & D_{23} \\ D_{13} & D_{23} & 1 \end{pmatrix}\begin{pmatrix} \sin\beta\cos\alpha \\ \sin\beta\sin\alpha \\ \cos\beta \end{pmatrix} = 1. \tag{H18}$$

Taking $\alpha = \pi/4, \beta = \pi/2$, we have

$$(\sqrt{2}/2 \quad \sqrt{2}/2 \quad 0)\begin{pmatrix} 1 & D_{12} & D_{13} \\ D_{12} & 1 & D_{23} \\ D_{13} & D_{23} & 1 \end{pmatrix}\begin{pmatrix} \sqrt{2}/2 \\ \sqrt{2}/2 \\ 0 \end{pmatrix} = 1 + D_{12} = 1. \tag{H19}$$

Thus we have

$$D_{12} = D_{21} = 0. \tag{H20}$$

Similarly, taking $\alpha = \pi/2, \beta = \pi/4$ and $\alpha = 0, \beta = \pi/4$ respectively, we have

$$D_{23} = D_{32} = 0, D_{13} = D_{31} = 0. \tag{H21}$$

Combining Eqs. (H16), (H17), (H20) and (H21), we have

$$D = (B^{-1})^T B^{-1} = \begin{pmatrix} 1 & 0 & 0 \\ 0 & 1 & 0 \\ 0 & 0 & 1 \end{pmatrix} = I. \tag{H22}$$

Thus we have

$$(B)^T B = \begin{pmatrix} 1 & 0 & 0 \\ 0 & 1 & 0 \\ 0 & 0 & 1 \end{pmatrix}. \tag{H23}$$

Combining Eqs. (H23), (H7-a, b, c) and (H9), $\hat{\mathbf{k}}_1$, $\hat{\mathbf{k}}_2$ and $\hat{\mathbf{k}}_3$ are three mutually perpendicular directions.

Therefore, the proposition is proved.

# Part I

**Proposition:**

For photons, if two momentum directions are perpendicular to each other, the corresponding spin planes are orthogonal, and vice versa.

**Proof:**

Suppose the two momentum directions $\hat{\mathbf{k}}_1$ and $\hat{\mathbf{k}}_2$ described by $\alpha_1, \beta_1$ and $\alpha_2, \beta_2$ are perpendicular to each other. According to the Eq. (H9) in Part H, we have

$$\sin\beta_1 \sin\beta_2 \cos\alpha_1 \cos\alpha_2 + \sin\beta_1 \sin\beta_2 \sin\alpha_1 \sin\alpha_2 + \cos\beta_1 \cos\beta_2 = 0. \tag{I1}$$

Below we prove that the two spin planes determined by $\left|S_{\hat{\mathbf{k}}_1} = +\hbar\right\rangle, \left|S_{\hat{\mathbf{k}}_1} = -\hbar\right\rangle$ and $\left|S_{\hat{\mathbf{k}}_2} = +\hbar\right\rangle, \left|S_{\hat{\mathbf{k}}_2} = -\hbar\right\rangle$ are orthogonal. Let $|S\rangle = \left|S_{\hat{\mathbf{k}}_2} = +\hbar\right\rangle + C\left|S_{\hat{\mathbf{k}}_2} = -\hbar\right\rangle$, where $C$ is a complex constant. Utilizing the Eqs. (1a, b), we have

$$\begin{aligned}|S\rangle &= \left|S_{\hat{\mathbf{k}}_2} = +\hbar\right\rangle + C\left|S_{\hat{\mathbf{k}}_2} = -\hbar\right\rangle \\ &= \frac{1}{2}e^{-i\alpha_2}[(1+\cos\beta_2) + C(1-\cos\beta_2)]|a\rangle + \frac{1}{\sqrt{2}}\sin\beta_2(1-C)|b\rangle \\ &\quad + \frac{1}{2}e^{i\alpha_2}[(1-\cos\beta_2) + C(1+\cos\beta_2)]|c\rangle.\end{aligned} \tag{I2}$$

Therefore,

$$\langle S_{\hat{\mathbf{k}}_1} = +\hbar | S \rangle = \frac{1}{4} e^{i(\alpha_1 - \alpha_2)} (1 + \cos \beta_1)[(1 + \cos \beta_2) + C(1 - \cos \beta_2)]$$
$$+ \frac{1}{2} \sin \beta_1 \sin \beta_2 (1 - C) + \frac{1}{4} e^{i(\alpha_2 - \alpha_1)} (1 - \cos \beta_1)[(1 - \cos \beta_2) + C(1 + \cos \beta_2)]. \quad (I3)$$

$$\langle S_{\hat{\mathbf{k}}_1} = -\hbar | S \rangle = \frac{1}{4} e^{i(\alpha_1 - \alpha_2)} (1 - \cos \beta_1)[(1 + \cos \beta_2) + C(1 - \cos \beta_2)]$$
$$- \frac{1}{2} \sin \beta_1 \sin \beta_2 (1 - C) + \frac{1}{4} e^{i(\alpha_2 - \alpha_1)} (1 + \cos \beta_1)[(1 - \cos \beta_2) + C(1 + \cos \beta_2)]. \quad (I4)$$

Below we prove that as long as the Eq. (I1) holds, then we can always find a $C$ satisfying both the equations $\langle S_{\hat{\mathbf{k}}_1} = +\hbar | S \rangle = 0$ and $\langle S_{\hat{\mathbf{k}}_1} = -\hbar | S \rangle = 0$, that is, making

$$\frac{1}{4} e^{i(\alpha_1 - \alpha_2)} (1 + \cos \beta_1)[(1 + \cos \beta_2) + C(1 - \cos \beta_2)] + \frac{1}{2} \sin \beta_1 \sin \beta_2 (1 - C)$$
$$+ \frac{1}{4} e^{i(\alpha_2 - \alpha_1)} (1 - \cos \beta_1)[(1 - \cos \beta_2) + C(1 + \cos \beta_2)] = 0, \quad (I5)$$

$$\frac{1}{4} e^{i(\alpha_1 - \alpha_2)} (1 - \cos \beta_1)[(1 + \cos \beta_2) + C(1 - \cos \beta_2)] - \frac{1}{2} \sin \beta_1 \sin \beta_2 (1 - C)$$
$$+ \frac{1}{4} e^{i(\alpha_2 - \alpha_1)} (1 + \cos \beta_1)[(1 - \cos \beta_2) + C(1 + \cos \beta_2)] = 0, \quad (I6)$$

hold. According to Eqs. (I5) and (I6), we have

$$e^{i(\alpha_1 - \alpha_2)}(1 + \cos \beta_1)(1 + \cos \beta_2) + 2 \sin \beta_1 \sin \beta_2 + e^{i(\alpha_2 - \alpha_1)}(1 - \cos \beta_1)(1 - \cos \beta_2)$$
$$+ C[e^{i(\alpha_1 - \alpha_2)}(1 + \cos \beta_1)(1 - \cos \beta_2) - 2 \sin \beta_1 \sin \beta_2 + e^{i(\alpha_2 - \alpha_1)}(1 - \cos \beta_1)(1 + \cos \beta_2)] \quad (I7)$$
$$= 0,$$

$$e^{i(\alpha_1 - \alpha_2)}(1 - \cos \beta_1)(1 + \cos \beta_2) - 2 \sin \beta_1 \sin \beta_2 + e^{i(\alpha_2 - \alpha_1)}(1 + \cos \beta_1)(1 - \cos \beta_2)$$
$$+ C[e^{i(\alpha_1 - \alpha_2)}(1 - \cos \beta_1)(1 - \cos \beta_2) + 2 \sin \beta_1 \sin \beta_2 + e^{i(\alpha_2 - \alpha_1)}(1 + \cos \beta_1)(1 + \cos \beta_2)] \quad (I8)$$
$$= 0.$$

For Eqs. (I7) and (I8), the condition for the existence of solution is

$$[e^{i(\alpha_1 - \alpha_2)}(1 + \cos \beta_1)(1 + \cos \beta_2) + 2 \sin \beta_1 \sin \beta_2 + e^{i(\alpha_2 - \alpha_1)}(1 - \cos \beta_1)(1 - \cos \beta_2)]$$
$$\cdot [e^{i(\alpha_1 - \alpha_2)}(1 - \cos \beta_1)(1 - \cos \beta_2) + 2 \sin \beta_1 \sin \beta_2 + e^{i(\alpha_2 - \alpha_1)}(1 + \cos \beta_1)(1 + \cos \beta_2)]$$
$$= [e^{i(\alpha_1 - \alpha_2)}(1 + \cos \beta_1)(1 - \cos \beta_2) - 2 \sin \beta_1 \sin \beta_2 + e^{i(\alpha_2 - \alpha_1)}(1 - \cos \beta_1)(1 + \cos \beta_2)] \quad (I9)$$
$$\cdot [e^{i(\alpha_1 - \alpha_2)}(1 - \cos \beta_1)(1 + \cos \beta_2) - 2 \sin \beta_1 \sin \beta_2 + e^{i(\alpha_2 - \alpha_1)}(1 + \cos \beta_1)(1 - \cos \beta_2)].$$

Namely,

$$16 \cos \beta_1 \cos \beta_2 + 8 e^{i(\alpha_1 - \alpha_2)} \sin \beta_1 \sin \beta_2 + 8 e^{i(\alpha_2 - \alpha_1)} \sin \beta_1 \sin \beta_2 = 0. \quad (I10)$$

Namely,

$$\cos \beta_1 \cos \beta_2 + \sin \beta_1 \sin \beta_2 \cos \alpha_1 \cos \alpha_2 + \sin \beta_1 \sin \beta_2 \sin \alpha_1 \sin \alpha_2 = 0. \quad (I11)$$

Obviously, the Eq. (I11) is just the Eq. (I1). Therefore, if the Eq. (I1) holds, we can always find a constant satisfying the Eqs. (I5) and (I6). Therefore, if two momentum directions are perpendicular to each other, the corresponding spin planes are orthogonal.

Below we prove the converse proposition. If the two spin planes determined by $\left|S_{\hat{\mathbf{k}}_1}=+\hbar\right\rangle,\left|S_{\hat{\mathbf{k}}_1}=-\hbar\right\rangle$ and $\left|S_{\hat{\mathbf{k}}_2}=+\hbar\right\rangle,\left|S_{\hat{\mathbf{k}}_2}=-\hbar\right\rangle$ are orthogonal, we can find a complex constant $C$ satisfying the Eqs. (I5) and (I6). Therefore the Eq. (I11), and namely, Eq. (I1) holds. Thus, two momentum directions $\hat{\mathbf{k}}_1$ and $\hat{\mathbf{k}}_2$ described by $\alpha_1, \beta_1$ and $\alpha_2, \beta_2$ respectively are perpendicular to each other.

Therefore, the proposition is proved.

## Part J

**Proposition:**

If

$$\begin{pmatrix} \hat{\mathbf{k}}_{x'} \\ \hat{\mathbf{k}}_{y'} \\ \hat{\mathbf{k}}_{z'} \end{pmatrix} = M \begin{pmatrix} \hat{\mathbf{k}}_x \\ \hat{\mathbf{k}}_y \\ \hat{\mathbf{k}}_z \end{pmatrix}, \tag{J1}$$

$$\begin{pmatrix} |x'\rangle \\ |y'\rangle \\ |z'\rangle \end{pmatrix} = N \begin{pmatrix} |x\rangle \\ |y\rangle \\ |z\rangle \end{pmatrix}, \tag{J2}$$

$$\hat{\mathbf{k}}_{x'}=f(|y'\rangle,|z'\rangle) , \hat{\mathbf{k}}_{y'}=f(|z'\rangle,|x'\rangle) , \hat{\mathbf{k}}_{z'} = f(|x'\rangle,|y'\rangle), \tag{J3}$$

where $M$ and $N$ are both 3×3 real orthogonal and $\det M = \det N = 1$, then

$$M = N. \tag{J4}$$

**Proof:**

Denote the elements of $M, N$ as $M_{ij}, N_{ij} (i, j = 1, 2, 3)$ respectively. According to Eq. (J1), we have

$$\hat{\mathbf{k}}_{x'} = M_{11}\hat{\mathbf{k}}_x + M_{12}\hat{\mathbf{k}}_y + M_{13}\hat{\mathbf{k}}_z. \tag{J5}$$

According to the Eq. (J2), we have

$$|y'\rangle = N_{21}|x\rangle + N_{22}|y\rangle + N_{23}|z\rangle, \tag{J6}$$

$$|z'\rangle = N_{31}|x\rangle + N_{32}|y\rangle + N_{33}|z\rangle. \tag{J7}$$

Substitute the Eqs. (J5), (J6) and (J7) into the first one of Eqs. (J3). We have

$$M_{11}\hat{\mathbf{k}}_{\mathbf{x}} + M_{12}\hat{\mathbf{k}}_{\mathbf{y}} + M_{13}\hat{\mathbf{k}}_{\mathbf{z}}$$
$$= f(N_{21}|x\rangle + N_{22}|y\rangle + N_{23}|z\rangle, N_{31}|x\rangle + N_{32}|y\rangle + N_{33}|z\rangle). \tag{J8}$$

According to the Eq. (23) in the main text, we have

$$M_{11}\hat{\mathbf{k}}_{\mathbf{x}} + M_{12}\hat{\mathbf{k}}_{\mathbf{y}} + M_{13}\hat{\mathbf{k}}_{\mathbf{z}}$$
$$= (N_{22}N_{33} - N_{23}N_{32})\hat{\mathbf{k}}_{\mathbf{x}} + (N_{23}N_{31} - N_{21}N_{33})\hat{\mathbf{k}}_{\mathbf{y}} + (N_{21}N_{32} - N_{22}N_{31})\hat{\mathbf{k}}_{\mathbf{z}}. \tag{J9}$$

Thus, we have

$$M_{11} = N_{22}N_{33} - N_{23}N_{32}, \quad M_{12} = N_{23}N_{31} - N_{21}N_{33}, \quad M_{13} = N_{21}N_{32} - N_{22}N_{31}. \tag{J10}$$

Similarly, we can obtain

$$M_{21} = N_{32}N_{13} - N_{33}N_{12}, \quad M_{22} = N_{33}N_{11} - N_{31}N_{13}, \quad M_{23} = N_{31}N_{12} - N_{32}N_{11},$$
$$M_{31} = N_{12}N_{23} - N_{13}N_{22}, \quad M_{32} = N_{13}N_{21} - N_{11}N_{23}, \quad M_{33} = N_{11}N_{22} - N_{12}N_{21}. \tag{J11}$$

Because $N$ is a 3×3 real matrix, and $\det N = 1$, it can be expressed with Euler angles as

$$N = Z'(\varphi)Y'(\theta)Z(\psi)$$
$$= \begin{pmatrix} \cos\varphi & \sin\varphi & 0 \\ -\sin\varphi & \cos\varphi & 0 \\ 0 & 0 & 1 \end{pmatrix} \begin{pmatrix} \cos\theta & 0 & -\sin\theta \\ 0 & 1 & 0 \\ \sin\theta & 0 & \cos\theta \end{pmatrix} \begin{pmatrix} \cos\psi & \sin\psi & 0 \\ -\sin\psi & \cos\psi & 0 \\ 0 & 0 & 1 \end{pmatrix}$$
$$= \begin{pmatrix} \cos\varphi\cos\theta\cos\psi - \sin\varphi\sin\psi & \cos\varphi\cos\theta\sin\psi + \sin\varphi\cos\psi & -\cos\varphi\sin\theta \\ -\sin\varphi\cos\theta\cos\psi - \cos\varphi\sin\psi & -\sin\varphi\cos\theta\sin\psi + \cos\varphi\cos\psi & \sin\varphi\sin\theta \\ \sin\theta\cos\psi & \sin\theta\sin\psi & \cos\theta \end{pmatrix}.$$
$$\tag{J12}$$

Using the expressions of matrix elements in Eq. (J12), we can obtain through direct calculations that

$$N_{22}N_{33} - N_{23}N_{32}$$
$$= (-\sin\varphi\cos\theta\sin\psi + \cos\varphi\cos\psi)\cos\theta - \sin\varphi\sin\theta\sin\theta\sin\psi$$
$$= \cos\varphi\cos\theta\cos\psi - \sin\varphi\sin\psi$$
$$= N_{11}. \tag{J13}$$

Thus, according to Eq. (J10), we have

$$M_{11} = N_{11}. \tag{J14}$$

Similarly, we can obtain through direct calculations that for all $i, j = 1, 2, 3$,

$$M_{ij} = N_{ij}. \tag{J15}$$

Therefore, the proposition is proved.

## Part K

If the rotation in the momentum space equates to the rotation in the position space, then the proposition in Part J means that the base states $|x\rangle, |y\rangle, |z\rangle$ transform in the same way under the rotation in position space (represented by the matrix M). Here we use this to calculate the transformation of $|S_{\hat{z}} = +\hbar\rangle$, $|S_{\hat{z}} = 0\hbar\rangle$ and $|S_{\hat{z}} = -\hbar\rangle$ under space rotation.

According to the analysis in the main text, the base states $|a\rangle, |b\rangle, |c\rangle$ are $|S_{\hat{z}} = +\hbar\rangle$, $|S_{\hat{z}} = 0\hbar\rangle$ and $|S_{\hat{z}} = -\hbar\rangle$, respectively. According to the Eq. (J2) in Part J and the Eq. (9) in the main text, we have

$$|x'\rangle \equiv -\frac{1}{\sqrt{2}} e^{i\pi/4}(|S_{\hat{z}'} = +\hbar\rangle - |S_{\hat{z}'} = -\hbar\rangle) = N_{11}|x\rangle + N_{12}|y\rangle + N_{13}|z\rangle$$
$$= N_{11}(-\frac{1}{\sqrt{2}})e^{i\pi/4}(|S_{\hat{z}} = +\hbar\rangle - |S_{\hat{z}} = -\hbar\rangle) + N_{12}\frac{i}{\sqrt{2}}(|S_{\hat{z}} = +\hbar\rangle + |S_{\hat{z}} = -\hbar\rangle) \quad \text{(K1)}$$
$$+ N_{13}e^{i\pi/4}|S_{\hat{z}} = 0\hbar\rangle,$$

$$|y'\rangle \equiv \frac{i}{\sqrt{2}} e^{i\pi/4}(|S_{\hat{z}'} = +\hbar\rangle + |S_{\hat{z}'} = -\hbar\rangle) = N_{21}|x\rangle + N_{22}|y\rangle + N_{23}|z\rangle$$
$$= N_{21}(-\frac{1}{\sqrt{2}})e^{i\pi/4}(|S_{\hat{z}} = +\hbar\rangle - |S_{\hat{z}} = -\hbar\rangle) + N_{22}\frac{i}{\sqrt{2}}(|S_{\hat{z}} = +\hbar\rangle + |S_{\hat{z}} = -\hbar\rangle) \quad \text{(K2)}$$
$$+ N_{23}e^{i\pi/4}|S_{\hat{z}} = 0\hbar\rangle,$$

$$|z'\rangle \equiv e^{i\pi/4}|S_{\hat{z}'} = 0\hbar\rangle) = N_{31}|x\rangle + N_{32}|y\rangle + N_{33}|z\rangle$$
$$= N_{31}(-\frac{1}{\sqrt{2}})e^{i\pi/4}(|S_{\hat{z}} = +\hbar\rangle - |S_{\hat{z}} = -\hbar\rangle) + N_{32}\frac{i}{\sqrt{2}}(|S_{\hat{z}} = +\hbar\rangle + |S_{\hat{z}} = -\hbar\rangle) \quad \text{(K3)}$$
$$+ N_{33}e^{i\pi/4}|S_{\hat{z}} = 0\hbar\rangle.$$

Thus we have

$$|S_{\hat{z}'} = +\hbar\rangle = -\frac{1}{2}(-N_{11} + iN_{12} - iN_{21} - N_{22})|S_{\hat{z}} = +\hbar\rangle$$
$$-\frac{1}{2}(N_{11} + iN_{12} + iN_{21} - N_{22})|S_{\hat{z}} = -\hbar\rangle - \frac{1}{\sqrt{2}}(N_{13} + iN_{23})|S_{\hat{z}} = 0\hbar\rangle, \quad \text{(K4)}$$

$$|S_{\hat{z}'} = -\hbar\rangle = \frac{1}{2}(-N_{11} + iN_{12} + iN_{21} + N_{22})|S_{\hat{z}} = +\hbar\rangle$$
$$+\frac{1}{2}(N_{11} + iN_{12} - iN_{21} + N_{22})|S_{\hat{z}} = -\hbar\rangle + \frac{1}{\sqrt{2}}(N_{13} - iN_{23})|S_{\hat{z}} = 0\hbar\rangle, \quad \text{(K5)}$$

$$|S_{\hat{z}'} = 0\hbar\rangle = -\frac{1}{\sqrt{2}}(N_{31} - iN_{32})|S_{\hat{z}} = +\hbar\rangle$$
$$+\frac{1}{\sqrt{2}}(N_{31} + iN_{32})|S_{\hat{z}} = -\hbar\rangle + N_{33}|S_{\hat{z}} = 0\hbar\rangle. \quad \text{(K6)}$$

Substitute the expressions of the matrix elements $N_{ij}$ in Eq. (J12) in Part J, and we have

$$|S_{\hat{z}'} = +\hbar\rangle = -\frac{1}{2}(-\cos\varphi\cos\theta\cos\psi + \sin\varphi\sin\psi + i\cos\varphi\cos\theta\sin\psi + i\sin\varphi\cos\psi$$
$$+i\sin\varphi\cos\theta\cos\psi + i\cos\varphi\sin\psi + \sin\varphi\cos\theta\sin\psi - \cos\varphi\cos\psi)|S_{\hat{z}} = +\hbar\rangle$$
$$-\frac{1}{2}(\cos\varphi\cos\theta\cos\psi - \sin\varphi\sin\psi + i\cos\varphi\cos\theta\sin\psi + i\sin\varphi\cos\psi \tag{K7}$$
$$-i\sin\varphi\cos\theta\cos\psi - i\cos\varphi\sin\psi + \sin\varphi\cos\theta\sin\psi - \cos\varphi\cos\psi)|S_{\hat{z}} = -\hbar\rangle$$
$$-\frac{1}{\sqrt{2}}(-\cos\varphi\sin\theta + i\sin\varphi\sin\theta)|S_{\hat{z}} = 0\hbar\rangle$$
$$= \frac{1}{2}(1+\cos\theta)e^{-i\varphi}e^{-i\psi}|S_{\hat{z}} = +\hbar\rangle + \frac{1}{2}(1-\cos\theta)e^{-i\varphi}e^{i\psi}|S_{\hat{z}} = -\hbar\rangle + \frac{1}{\sqrt{2}}\sin\theta e^{-i\varphi}|S_{\hat{z}} = 0\hbar\rangle,$$

$$|S_{\hat{z}'} = -\hbar\rangle = \frac{1}{2}(-\cos\varphi\cos\theta\cos\psi + \sin\varphi\sin\psi + i\cos\varphi\cos\theta\sin\psi + i\sin\varphi\cos\psi$$
$$-i\sin\varphi\cos\theta\cos\psi - i\cos\varphi\sin\psi - \sin\varphi\cos\theta\sin\psi + \cos\varphi\cos\psi)|S_{\hat{z}} = +\hbar\rangle$$
$$+\frac{1}{2}(\cos\varphi\cos\theta\cos\psi - \sin\varphi\sin\psi + i\cos\varphi\cos\theta\sin\psi + i\sin\varphi\cos\psi \tag{K8}$$
$$+i\sin\varphi\cos\theta\cos\psi + i\cos\varphi\sin\psi - \sin\varphi\cos\theta\sin\psi + \cos\varphi\cos\psi)|S_{\hat{z}} = -\hbar\rangle$$
$$+\frac{1}{\sqrt{2}}(-\cos\varphi\sin\theta - i\sin\varphi\sin\theta)|S_{\hat{z}} = 0\hbar\rangle$$
$$= \frac{1}{2}(1-\cos\theta)e^{i\varphi}e^{-i\psi}|S_{\hat{z}} = +\hbar\rangle + \frac{1}{2}(1+\cos\theta)e^{i\varphi}e^{i\psi}|S_{\hat{z}} = -\hbar\rangle - \frac{1}{\sqrt{2}}\sin\theta e^{i\varphi}|S_{\hat{z}} = 0\hbar\rangle,$$

$$|S_{\hat{z}'} = 0\hbar\rangle = -\frac{1}{\sqrt{2}}(\sin\theta\cos\psi - i\sin\theta\sin\psi)|S_{\hat{z}} = +\hbar\rangle$$
$$+\frac{1}{\sqrt{2}}(\sin\theta\cos\psi + i\sin\theta\sin\psi)|S_{\hat{z}} = -\hbar\rangle + \cos\theta|S_{\hat{z}} = 0\hbar\rangle \tag{K9}$$
$$= -\frac{1}{\sqrt{2}}\sin\theta e^{-i\psi}|S_{\hat{z}} = +\hbar\rangle + \frac{1}{\sqrt{2}}\sin\theta e^{i\psi}|S_{\hat{z}} = -\hbar\rangle + \cos\theta|S_{\hat{z}} = 0\hbar\rangle.$$

Take $\varphi = 0$, and replace $\psi$ and $\theta$ with $\alpha$ and $\beta$ respectively. We have

$$|S_{\hat{z}'} = +\hbar\rangle = \frac{1}{2}(1+\cos\beta)e^{-i\alpha}|S_{\hat{z}} = +\hbar\rangle + \frac{1}{2}(1-\cos\beta)e^{i\alpha}|S_{\hat{z}} = -\hbar\rangle + \frac{1}{\sqrt{2}}\sin\beta|S_{\hat{z}} = 0\hbar\rangle, \quad (K10)$$

$$|S_{\hat{z}'} = -\hbar\rangle = \frac{1}{2}(1-\cos\beta)e^{-i\alpha}|S_{\hat{z}} = +\hbar\rangle + \frac{1}{2}(1+\cos\beta)e^{i\alpha}|S_{\hat{z}} = -\hbar\rangle - \frac{1}{\sqrt{2}}\sin\beta|S_{\hat{z}} = 0\hbar\rangle, \quad (K11)$$

$$|S_{\hat{z}'} = 0\hbar\rangle = -\frac{1}{\sqrt{2}}\sin\beta e^{-i\alpha}|S_{\hat{z}} = +\hbar\rangle + \frac{1}{\sqrt{2}}\sin\beta e^{i\alpha}|S_{\hat{z}} = -\hbar\rangle + \cos\beta|S_{\hat{z}} = 0\hbar\rangle. \tag{K12}$$

The Eqs. (K10), (K11) and (K12) are the just the well-known transformation formulas of the base states $|S_{\hat{z}} = +\hbar\rangle$, $|S_{\hat{z}} = 0\hbar\rangle$ and $|S_{\hat{z}} = -\hbar\rangle$, for a spin-1 particle under space rotation. (see J. J. Sakurai and J. Napolitano, *Modern Quantum Mechanics*, Chapter 3 (China Edition, Beijing World Publishing Corporation, 2013) ).

# Part L

**Proposition**:

Both of the spin planes $(|a\rangle, |b\rangle)$ and $(|b\rangle, |c\rangle)$ can not be obtained from the following two base states by choosing the values of $\alpha$ and $\beta$,

$$\left|S_{\hat{\mathbf{k}}} = +\hbar\right\rangle = \frac{1}{2}(1+\cos\beta)e^{-i\alpha}|a\rangle + \frac{1}{\sqrt{2}}\sin\beta|b\rangle + \frac{1}{2}(1-\cos\beta)e^{i\alpha}|c\rangle, \quad \text{(L1)}$$

$$\left|S_{\hat{\mathbf{k}}} = -\hbar\right\rangle = \frac{1}{2}(1-\cos\beta)e^{-i\alpha}|a\rangle - \frac{1}{\sqrt{2}}\sin\beta|b\rangle + \frac{1}{2}(1+\cos\beta)e^{i\alpha}|c\rangle. \quad \text{(L2)}$$

**Proof**:

Consider the spin plane $(|a\rangle, |b\rangle)$. If $\alpha_0$ and $\beta_0$ are present such that the spin plane determined by $\left|S_{\hat{\mathbf{k}}} = +\hbar\right\rangle_{\alpha=\alpha_0, \beta=\beta_0}$ and $\left|S_{\hat{\mathbf{k}}} = -\hbar\right\rangle_{\alpha=\alpha_0, \beta=\beta_0}$ is the same spin plane determined by $|a\rangle$ and $|b\rangle$, then the following equations hold

$$m_{11}\left|S_{\hat{\mathbf{k}}} = +\hbar\right\rangle_{\alpha=\alpha_0, \beta=\beta_0} + m_{12}\left|S_{\hat{\mathbf{k}}} = -\hbar\right\rangle_{\alpha=\alpha_0, \beta=\beta_0} = |a\rangle, \quad \text{(L3)}$$

$$m_{21}\left|S_{\hat{\mathbf{k}}} = +\hbar\right\rangle_{\alpha=\alpha_0, \beta=\beta_0} + m_{22}\left|S_{\hat{\mathbf{k}}} = -\hbar\right\rangle_{\alpha=\alpha_0, \beta=\beta_0} = |b\rangle. \quad \text{(L4)}$$

Substitute the Eqs. (L1) and (L2), and we have

$$\frac{1}{2}e^{-i\alpha_0}[m_{11}(1+\cos\beta_0) + m_{12}(1-\cos\beta_0)]|a\rangle + \frac{1}{\sqrt{2}}\sin\beta_0[m_{11} - m_{12}]|b\rangle$$
$$+ \frac{1}{2}e^{i\alpha_0}[m_{11}(1-\cos\beta_0) + m_{12}(1+\cos\beta_0)]|c\rangle = |a\rangle. \quad \text{(L5)}$$

$$\frac{1}{2}e^{-i\alpha_0}[m_{21}(1+\cos\beta_0) + m_{22}(1-\cos\beta_0)]|a\rangle + \frac{1}{\sqrt{2}}\sin\beta_0[m_{21} - m_{22}]|b\rangle$$
$$+ \frac{1}{2}e^{i\alpha_0}[m_{21}(1-\cos\beta_0) + m_{22}(1+\cos\beta_0)]|c\rangle = |b\rangle. \quad \text{(L6)}$$

Thus we have

$$\frac{1}{2}e^{-i\alpha_0}[m_{11}(1+\cos\beta_0) + m_{12}(1-\cos\beta_0)] = 1, \quad \text{(L7-a)}$$

$$\frac{1}{\sqrt{2}}\sin\beta_0[m_{11} - m_{12}] = 0, \quad \text{(L7-b)}$$

$$\frac{1}{2}e^{-i\alpha_0}[m_{21}(1+\cos\beta_0) + m_{22}(1-\cos\beta_0)] = 0, \quad \text{(L7-c)}$$

$$\frac{1}{2}e^{-i\alpha_0}[m_{21}(1+\cos\beta_0) + m_{22}(1-\cos\beta_0)] = 0, \quad \text{(L7-d)}$$

$$\frac{1}{\sqrt{2}}\sin\beta_0[m_{21} - m_{22}] = 1, \quad \text{(L7-e)}$$

$$\frac{1}{2}e^{i\alpha_0}[m_{21}(1-\cos\beta_0)+m_{22}(1+\cos\beta_0)]=0. \tag{L7-f}$$

According to Eq. (L7-e), we have $\sin\beta_0 \neq 0$. Therefore, according to Eq. (L7-b), we have $m_{11}=m_{12}$. Substituting it into Eq. (L7-a, c), we have

$$m_{11}e^{-i\alpha_0}=1, \tag{L8-a}$$

$$m_{11}e^{i\alpha_0}=0. \tag{L8-b}$$

According to Eq. (L8-b), we have $m_{11}=0$. Thus, Eq. (L8-a) can not hold. Therefore, Eqs. (L3) and (L4) can not hold. Thus, there is no $\alpha_0$ and $\beta_0$ that make the spin plane determined by $\left|S_{\hat{\mathbf{k}}}=+\hbar\right\rangle_{\alpha=\alpha_0,\beta=\beta_0}$ and $\left|S_{\hat{\mathbf{k}}}=-\hbar\right\rangle_{\alpha=\alpha_0,\beta=\beta_0}$ is the spin plane determined by $|a\rangle$ and $|b\rangle$. Similarly, it can be proved that there is no $\alpha_0$ and $\beta_0$ that make the spin plane determined by $\left|S_{\hat{\mathbf{k}}}=+\hbar\right\rangle_{\alpha=\alpha_0,\beta=\beta_0}$ and $\left|S_{\hat{\mathbf{k}}}=-\hbar\right\rangle_{\alpha=\alpha_0,\beta=\beta_0}$ is the spin plane determined by $|b\rangle$ and $|c\rangle$.

Therefore, the proposition is proved.

## Part M

**Proposition:**

For any spin base states $|a'\rangle$, $|b'\rangle$ and $|c'\rangle$, the equation

$$\begin{aligned}&f(a_{11}|a'\rangle+a_{12}|b'\rangle+a_{13}|c'\rangle,a_{21}|a'\rangle+a_{22}|b'\rangle+a_{23}|c'\rangle)\\&=(a_{11}a_{22}-a_{12}a_{21})f(|a'\rangle,|b'\rangle)+(a_{12}a_{23}-a_{13}a_{22})f(|b'\rangle,|c'\rangle)+(a_{13}a_{21}-a_{11}a_{23})f(|c'\rangle,|a'\rangle),\end{aligned} \tag{M1}$$

holds if $f(|a'\rangle,|b'\rangle)$, $f(|b'\rangle,|c'\rangle)$ and $f(|c'\rangle,|a'\rangle)$ are understood as linear combinations of $f(|x\rangle,|y\rangle)$, $f(|y\rangle,|z\rangle)$ and $f(|z\rangle,|x\rangle)$ in the sense of Eq. (23) in the main text.

**Proof:**

Let

$$|a'\rangle=A_{11}|x\rangle+A_{12}|y\rangle+A_{13}|z\rangle, \tag{M2-a}$$

$$|b'\rangle=A_{21}|x\rangle+A_{22}|y\rangle+A_{23}|z\rangle, \tag{M2-b}$$

$$|c'\rangle=A_{31}|x\rangle+A_{32}|y\rangle+A_{33}|z\rangle. \tag{M2-c}$$

Substituting them into the left sides of Eq. (M1), and expanding it according to Eq. (23) in the main text, we have

$$f(a_{11}|a'\rangle + a_{12}|b'\rangle + a_{13}|c'\rangle, a_{21}|a'\rangle + a_{22}|b'\rangle + a_{23}|c'\rangle)$$
$$= f((a_{11}A_{11} + a_{12}A_{21} + a_{13}A_{31})|x\rangle + (a_{11}A_{12} + a_{12}A_{22} + a_{13}A_{32})|y\rangle + (a_{11}A_{13} + a_{12}A_{23} + a_{13}A_{33})|z\rangle$$
$$(a_{21}A_{11} + a_{22}A_{21} + a_{23}A_{31})|x\rangle + (a_{21}A_{12} + a_{22}A_{22} + a_{23}A_{32})|y\rangle + (a_{21}A_{13} + a_{22}A_{23} + a_{23}A_{33})|z\rangle)$$
$$= [(a_{11}A_{11} + a_{12}A_{21} + a_{13}A_{31})(a_{21}A_{12} + a_{22}A_{22} + a_{23}A_{32})$$
$$-(a_{11}A_{12} + a_{12}A_{22} + a_{13}A_{32})(a_{21}A_{11} + a_{22}A_{21} + a_{23}A_{31})]f(|x\rangle, |y\rangle)$$
$$+[(a_{11}A_{12} + a_{12}A_{22} + a_{13}A_{32})(a_{21}A_{13} + a_{22}A_{23} + a_{23}A_{33})$$
$$-(a_{11}A_{13} + a_{12}A_{23} + a_{13}A_{33})(a_{21}A_{12} + a_{22}A_{22} + a_{23}A_{32})]f(|y\rangle, |z\rangle)$$
$$+[(a_{11}A_{13} + a_{12}A_{23} + a_{13}A_{33})(a_{21}A_{11} + a_{22}A_{21} + a_{23}A_{31})$$
$$-(a_{11}A_{11} + a_{12}A_{21} + a_{13}A_{31})(a_{21}A_{13} + a_{22}A_{23} + a_{23}A_{33})]f(|z\rangle, |x\rangle)$$
$$= [(a_{11}a_{22} - a_{12}a_{21})(A_{11}A_{22} - A_{12}A_{21}) + (a_{11}a_{23} - a_{13}a_{21})(A_{11}A_{32} - A_{12}A_{31})$$
$$+(a_{12}a_{23} - a_{13}a_{22})(A_{21}A_{32} - A_{22}A_{31})]f(|x\rangle, |y\rangle)$$
$$+[(a_{11}a_{22} - a_{12}a_{21})(A_{12}A_{23} - A_{13}A_{22}) + (a_{11}a_{23} - a_{13}a_{21})(A_{12}A_{33} - A_{13}A_{32})$$
$$+(a_{12}a_{23} - a_{13}a_{22})(A_{22}A_{33} - A_{23}A_{32})]f(|y\rangle, |z\rangle)$$
$$+[(a_{11}a_{22} - a_{12}a_{21})(A_{13}A_{21} - A_{11}A_{23}) + (a_{11}a_{23} - a_{13}a_{21})(A_{13}A_{31} - A_{11}A_{33})$$
$$+(a_{12}a_{23} - a_{13}a_{22})(A_{23}A_{31} - A_{21}A_{33})]f(|z\rangle, |x\rangle). \qquad (M3)$$

Similarly, substituting Eqs. (M2-a, b, c) into the right side of Eq. (M1), and expanding it according to Eq. (23) in the main text, we have

$$(a_{11}a_{22} - a_{12}a_{21})f(|a'\rangle, |b'\rangle) + (a_{12}a_{23} - a_{13}a_{22})f(|b'\rangle, |c'\rangle) + (a_{13}a_{21} - a_{11}a_{23})f(|c'\rangle, |a'\rangle)$$
$$= (a_{11}a_{22} - a_{12}a_{21})f(A_{11}|x\rangle + A_{12}|y\rangle + A_{13}|z\rangle, A_{21}|x\rangle + A_{22}|y\rangle + A_{23}|z\rangle)$$
$$+(a_{12}a_{23} - a_{13}a_{22})f(A_{21}|x\rangle + A_{22}|y\rangle + A_{23}|z\rangle, A_{31}|x\rangle + A_{32}|y\rangle + A_{33}|z\rangle)$$
$$+(a_{13}a_{21} - a_{11}a_{23})f(A_{31}|x\rangle + A_{32}|y\rangle + A_{33}|z\rangle, A_{11}|x\rangle + A_{12}|y\rangle + A_{13}|z\rangle)$$
$$= (a_{11}a_{22} - a_{12}a_{21})[(A_{11}A_{22} - A_{12}A_{21})f(|x\rangle, |y\rangle) + (A_{12}A_{23} - A_{13}A_{22})f(|y\rangle, |z\rangle)$$
$$+(A_{13}A_{21} - A_{11}A_{23})f(|z\rangle, |x\rangle)]$$
$$+(a_{12}a_{23} - a_{13}a_{22})[(A_{21}A_{32} - A_{22}A_{31})f(|x\rangle, |y\rangle) + (A_{22}A_{33} - A_{23}A_{32})f(|y\rangle, |z\rangle)$$
$$+(A_{23}A_{31} - A_{21}A_{33})f(|z\rangle, |x\rangle)]$$
$$+(a_{13}a_{21} - a_{11}a_{23})[(A_{31}A_{12} - A_{32}A_{11})f(|x\rangle, |y\rangle) + (A_{32}A_{13} - A_{33}A_{12})f(|y\rangle, |z\rangle)$$
$$+(A_{33}A_{11} - A_{31}A_{13})f(|z\rangle, |x\rangle)]$$
$$= [(a_{11}a_{22} - a_{12}a_{21})(A_{11}A_{22} - A_{12}A_{21}) + (a_{12}a_{23} - a_{13}a_{22})(A_{21}A_{32} - A_{22}A_{31})$$
$$+(a_{13}a_{21} - a_{11}a_{23})(A_{31}A_{12} - A_{32}A_{11})]f(|x\rangle, |y\rangle)$$
$$+[(a_{11}a_{22} - a_{12}a_{21})(A_{12}A_{23} - A_{13}A_{22}) + (a_{12}a_{23} - a_{13}a_{22})(A_{22}A_{33} - A_{23}A_{32})$$
$$+(a_{13}a_{21} - a_{11}a_{23})(A_{32}A_{13} - A_{33}A_{12})]f(|y\rangle, |z\rangle)$$
$$+[(a_{11}a_{22} - a_{12}a_{21})(A_{13}A_{21} - A_{11}A_{23}) + (a_{12}a_{23} - a_{13}a_{22})(A_{23}A_{31} - A_{21}A_{33})$$
$$+(a_{13}a_{21} - a_{11}a_{23})(A_{33}A_{11} - A_{31}A_{13})]f(|z\rangle, |x\rangle). \qquad (M4)$$

Obviously, the right sides of Eqs. (M3) and (M4) are equivalent.

Therefore, the proposition is proved.